# A novel generative reverse net assisted evolution algorithm for expensive-computational optimizations

**Yu Li[a, *], Hu Wang[a, **], Ziming Wen[a], Xin Wang[a]**

*a. State Key Laboratory of Advanced Design and Manufacturing for Vehicle Body, College of Mechanical and Vehicle Engineering, Hunan University, Changsha, 410082, PR China*

## Highlights

i. A Generative Reverse Net assisted Evolution Algorithm is proposed to accelerate expensive-simulation optimizations.

ii. A generation model, Residual Variational Auto-Encoder, is proposed.

iii. A novel method to filter out unreasonable generated cases is introduced.

iv. A Case-Diversity Rule to evaluate generation diversity for regression problems is proposed.

## Abstract

Simulation-based optimization is a useful method for practical design problems. However, it is difficult for complicated problems due to expensive-computational costs. A popular way to overcome this issue is to use a surrogate model to save the cost. Nevertheless, limited design parameters those are input to traditional surrogate models can difficultly represent the whole design problem, which might result in unexpected errors. In this study, physical cloud images from simulations are employed and attempted to construct the surrogate model. Simultaneously, based on the strong pattern recognition and generation abilities of deep learning models, a novel Generative Reverse Net assisted Evolution Algorithm (GRN-EA) is proposed for expensive-design problems. In this study, a numerical example of a Variable-Stiffness (VS) composite hole-plate is employed to obtain the optimal distribution of the curved fiber. Moreover, to evaluate the performance of GRN-EA in

[*] First author. *E-mail address*: liyu_hnu@hnu.edu.cn (Y. Li)

[**] Corresponding author. Tel.: +86 0731 88655012; fax: +86 0731 88822051.

*E-mail address:* wanghu@hnu.edu.cn (H. Wang)

practical engineering problems, a more complex sheet-forming case is optimized. According to the two problems, some expensive-simulations can be well handled in this study.

*Keywords*: Expensive simulation optimization; Deep neural network; Generative Reverse Net assisted Evolution Algorithm (GRN-EA)

## 1. Introduction

Evolution Algorithms (EAs) require a large number of function evaluations to converge to global-optimal or near-optimal solutions. This issue limits the capability of solving design problems relying on computationally expensive simulation models. Three popular ways [1] are commonly used to overcome such difficulty. The first one is fitness inheritance. For example, Bui [2] used the fitness inheritance to reduce the computational cost for six expensive-optimization problems. The second one is the fitness imitation [3, 4]. It is different from the fitness inheritance, and the fitness values of the offspring individuals are estimated from some selected representative individuals. The last one is to generate an approximation model called the surrogate model or metamodel [5]. The surrogate models are approximate models to study complex input-output relationships exhibited by another more complex model (physics-based model). Since the surrogate models can lead to the best performance and are the most efficient method, most researchers are concentrating on this approach. E.g., Shu [1] introduced an On-Line Variable-Fidelity Metamodel assisted Multi-Objective Genetic Algorithm (OLVFM-MOGA) approach and significantly reduced the computational cost with satisfied accuracy. Liu [6] and Xiao [7] used the surrogate models to optimize practical problems with low-costs. At the same time, Zadeh [8] presented surrogate-model based multi-objective multidisciplinary design optimization (MDO) architecture, which had high diversity and fast convergence. Moreover, Bucher [9] discussed strategies to obtain a suitable surrogate model, to assess its quality concerning prediction and to use the mode to do structural reliability analysis. Actually, traditional surrogate models, such as Radial Basis Function-High Dimensional Model Representation (RBF-HDMR) [10], Kriging (KG) [11], Least

Square Support Vector Regression (LSSVR) [12], etc., usually construct the mapping relationship from design parameters to the desired actual response (e.g., maximum/minimum or mean value). However, the limited parameters can difficultly represent the design problems perfectly, which might result in unexpected errors. On the other hand, physical cloud images contain more objective information, and can show the design problems intuitively. Accordingly, the physical cloud images are attempted to the surrogate model-based optimizations in this study.

Recently, with the explosive development of deep learning methods, many novel and efficient Deep Neural Network (DPNN) models have been introduced. The DPNNs can be classified into three categories, supervised learning, unsupervised learning and reinforcement learning. The supervised learning generates a function that maps inputs to desired outputs and usually do classifications and regressions. There are some typical supervised learning models, e.g., Back Propagation Neural Network (BPNN) [13], Convolutional Neural Network (CNN) [14, 15] and Recurrent Neural Network (RNN) [16]. Unsupervised learning models a set of inputs, liking clustering. It is popularly applied to generate, such as Generative Adversarial Network (GAN) [17] and Auto-Encoder (AE) [18]. Lastly, reinforcement learning [19] learns how to act when the environment provides feedbacks to guide the learning. Importantly, many models have been introduced to the optimizations, e.g., topology optimization [20, 21]. Albanesi [22] combined the GA with the DPNN to reduce the computational cost of the wind turbine blade optimizations. Simultaneously, Javier [23] presented a methodology by using the DPNNs to the designs and optimizations of one-way slabs. Moreover, Patnaik [24] employed the DPNNs and regression approximates to reanalysis and sensitivity analysis calculations at the sub-problem level. In addition, Wang [25] introduced the reinforcement learning to the surrogate models and proposed an automated meta-modeling framework.

In this study, aiming at expensive design and optimization problems, physical cloud images are employed to construct the models. Meanwhile, through strong pattern recognition and generation abilities of the DPNNs, a novel Generative Reverse Net assisted Evolution Algorithm (GRN-EA) is proposed to reduce the simulation cost and to accelerate optimizations for expensive-simulation problems.

The remainder of this paper is organized as follows. In Section 2, the architecture of the novel GRN-EA is introduced detailly. Then a numerical example of a variable-stiffness composite hole-plate is optimized by the GRN-EA to achieve the optimal distribution of the curved fiber in the following section. Simultaneously, to further evaluate the performance of the proposed method, the optimization of a more complex engineering application, a sheet forming problem, is also given. Ultimately, some perspective remarks are provided in Section 4.

## 2. Generative Reverse Network assisted Evolution Algorithm

In the simulation-based optimization processes, the most difficult part is the expensive-computational costs. As shown in Fig. 1, in the GRN-EA, the expensive simulation is represented by the GRN, which can largely reduce the simulation times for the optimization.

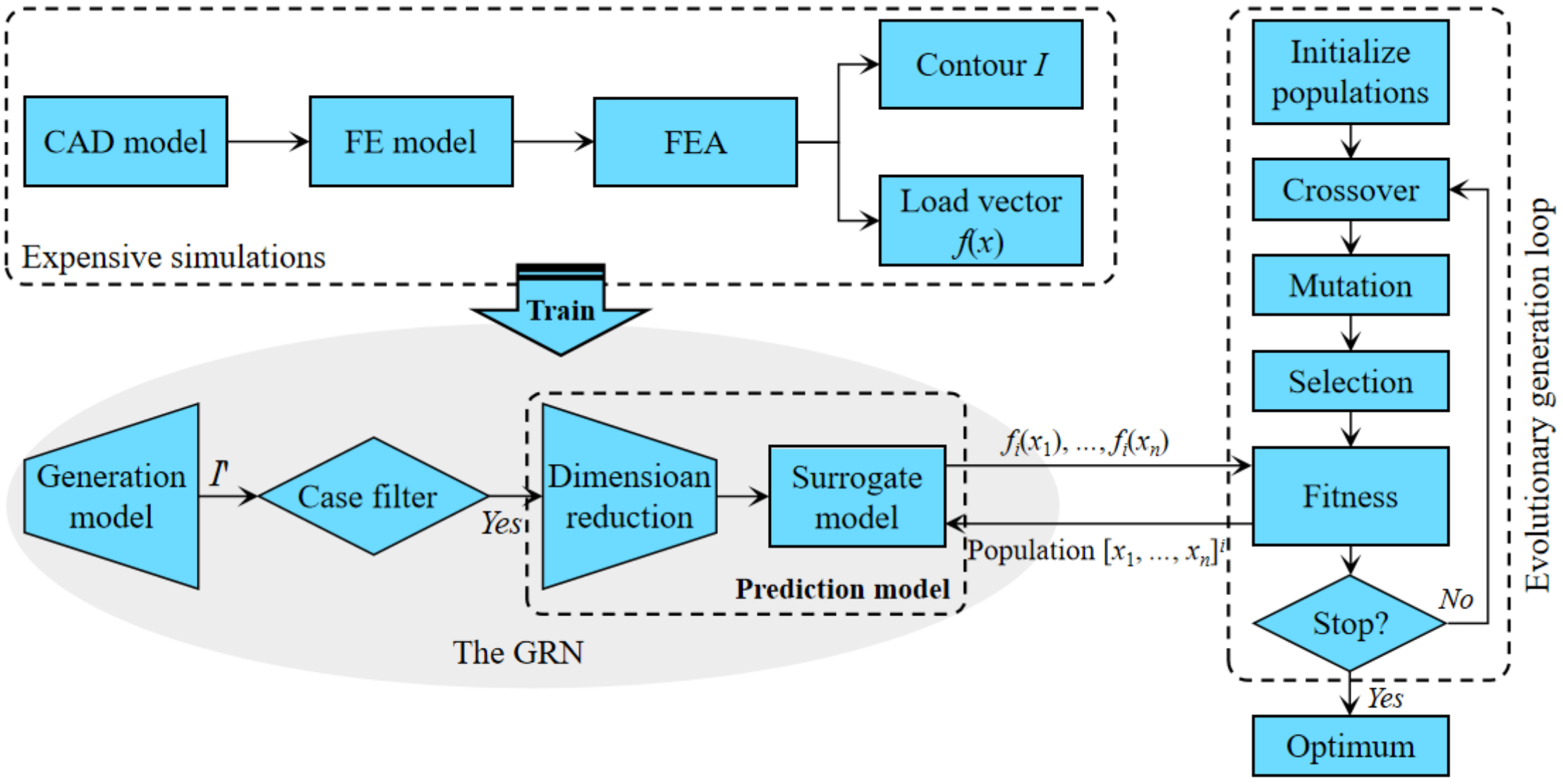

**Fig. 1.** The architecture of the GRN-EA method.

### 2.1. Generative Reverse Net

As shown in Fig. 1, the GRN mainly contains three parts, generation model that is used to generated simulated cases, prediction model to calculate the actual responses of the generated results, and new case filter to ensure that all generated results handled by the EA are reasonable.

2.1.1. Generation model

In this study, two popular generation models, Generative Adversarial Network (GAN) and Variational Auto-Encoder (VAE) [26], are considered. The GAN consists two functions, generator $G(z)$ that maps a sample (depending on a random or a Gaussian distribution) to the data distribution, and discriminator $D(x)$ that determines if an input belongs to the training data set. Mathematically, the training process can be described as

$$\min_{G}\max_{D} V(D,G) = \mathbb{E}_{\mathbf{x}\sim Pdata}\left[\log D(\mathbf{x})\right] + \mathbb{E}_{\mathbf{z}\sim Pz}\left[\log\left(1-D\left(G\left(\mathbf{z}\right)\right)\right)\right] \tag{1}$$

where $\mathbf{x}$ is the image from training samples $P_{data}$; and $\mathbf{z}$ is a noise vector sampled from distribution $P_z$.

The VAE is a manifold learning model. By maximization of the marginal likelihoods, as calculated in Eq. (2), the generation model can be constructed.

$$\log p_\theta(\mathbf{x}) = D_{KL}\left(q_\phi(\mathbf{z}\,|\,\mathbf{x})\,\|\,p_\theta(\mathbf{z}\,|\,\mathbf{x})\right) + \mathfrak{L}(\theta,\phi;\mathbf{x}) \tag{2}$$

where $q_\varphi(\mathbf{z}|\mathbf{x})$ is a recognition model as a probabilistic encoder; $p_\theta(\mathbf{x}|\mathbf{z})$ is a probabilistic decoder; and $\mathfrak{L}(\theta,\phi;\mathbf{x})$ is called the (variational) lower bound on the marginal likelihood.

Some comparisons of the GAN and VAE are discussed as follows.

i. The FE simulation is an expensive-computational process, and the number of training samples is limited.

ii. Because of the more complex structure of the GAN, more parameters need to be trained. Its computational costs of training are several times that of the VAE.

iii. The GAN uses a Gaussian or uniform distribution, rather than an assumed distribution, to approximate the real design space as much as possible in theory. However, as shown in Fig. 2, the dataset contains 400 samples from a sheet forming problem that is tested in Section 3.2, and the physical cloud images of all samples are compressed as 2-dimensional data ($x_1$, $x_2$) to be visualized. It can be found that the limited samples can hardly describe the design space, especially in the sparse red region. Thus, GAN needs more training samples to better describe the design space.

iv. VAE is a manifold learning model. As shown in Eq. (2), the first term, Kullback-Leibler (KL) divergence, helps the VAE generate new data to approximate the real design space based on one data $\mathbf{x}^i$ instead of the total set $p(\mathbf{x}^i)$. Thus, the dependence on the total dataset of the VAE is weaker. Furthermore, for the input being a random noise matrix **z**, the images in the manifold space are relevant to each other. As shown in Fig. 2, a surrounding space of a data $\mathbf{x}^i$ can be inferred by the VAE. Thus, the VAE has more advantages for a compact dataset, namely the distance between two data ($\mathbf{x}^i$ and $\mathbf{x}^k$) are small and the described space is clustered. Consequently, the VAE need less training samples and lower-computational costs.

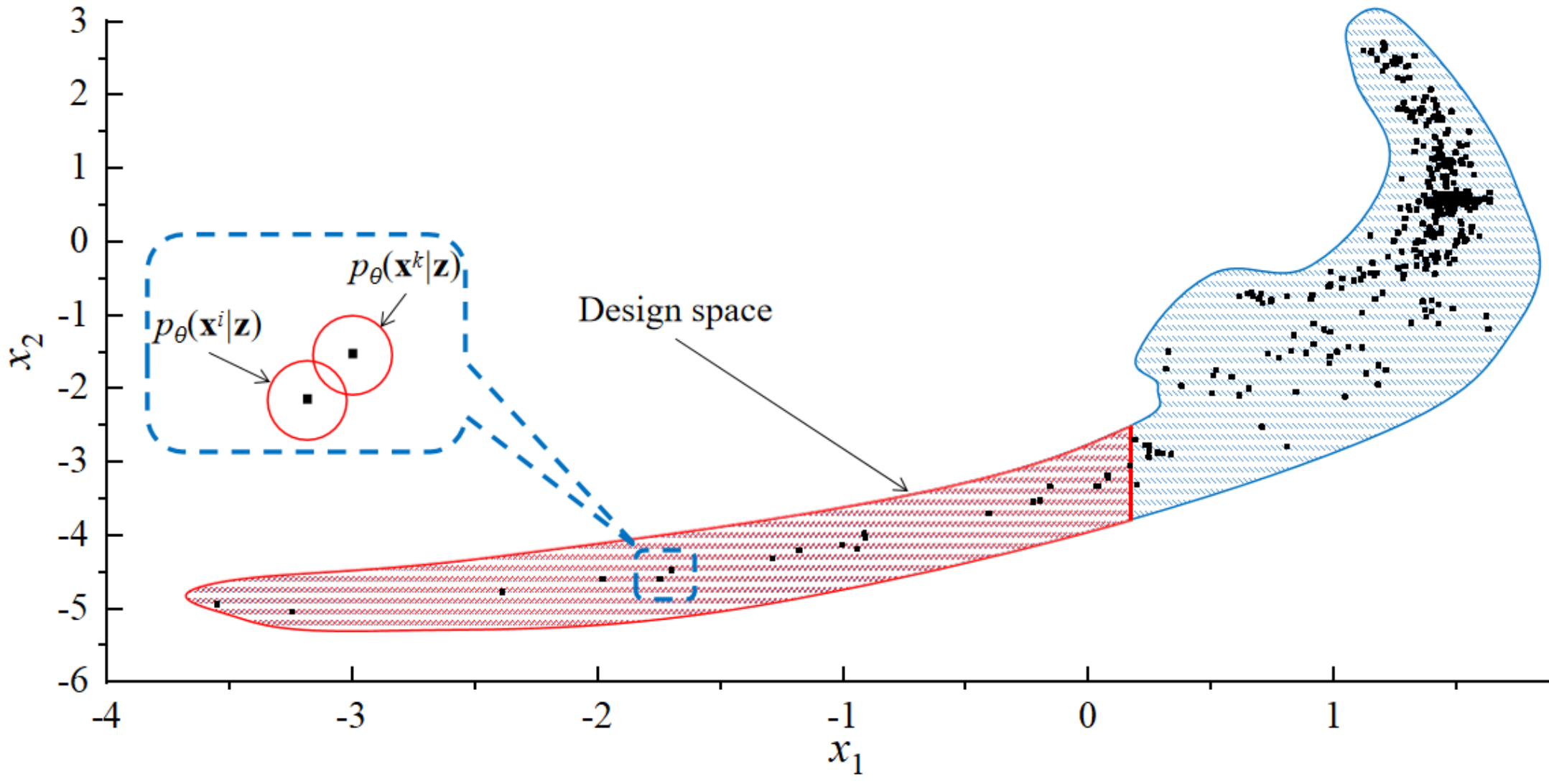

**Fig. 2.** Demonstration of the compact dataset distribution.

Consequently, the VAE is employed as the generation model. It is a convolutional-deconvolutional process. Additionally, because convolutional layers can extract low- or high-level features, the more the layers, the more obtained information. However, if the network is too deep, it is easily to be gradient explosion or disappearance. Hence, deeper neural networks are more difficult to be trained. Here, Residual block [27] is employed to the VAE, namely ResVAE. Through skip connections, the network with hundreds of layers can be trained.

As shown in Figs. 3 (a) – (b), $H(x)$ is the output from the 3rd layer of the Residual block.

$$H(x) = W_3\left(\sigma\left(W_2\left(\sigma\left(W_1 x\right)\right)\right)\right) \quad (3)$$

where $\sigma$ (·) represent Rectified Linear Unit (ReLU) function; and $W_i$ is each layer's weight matrix.

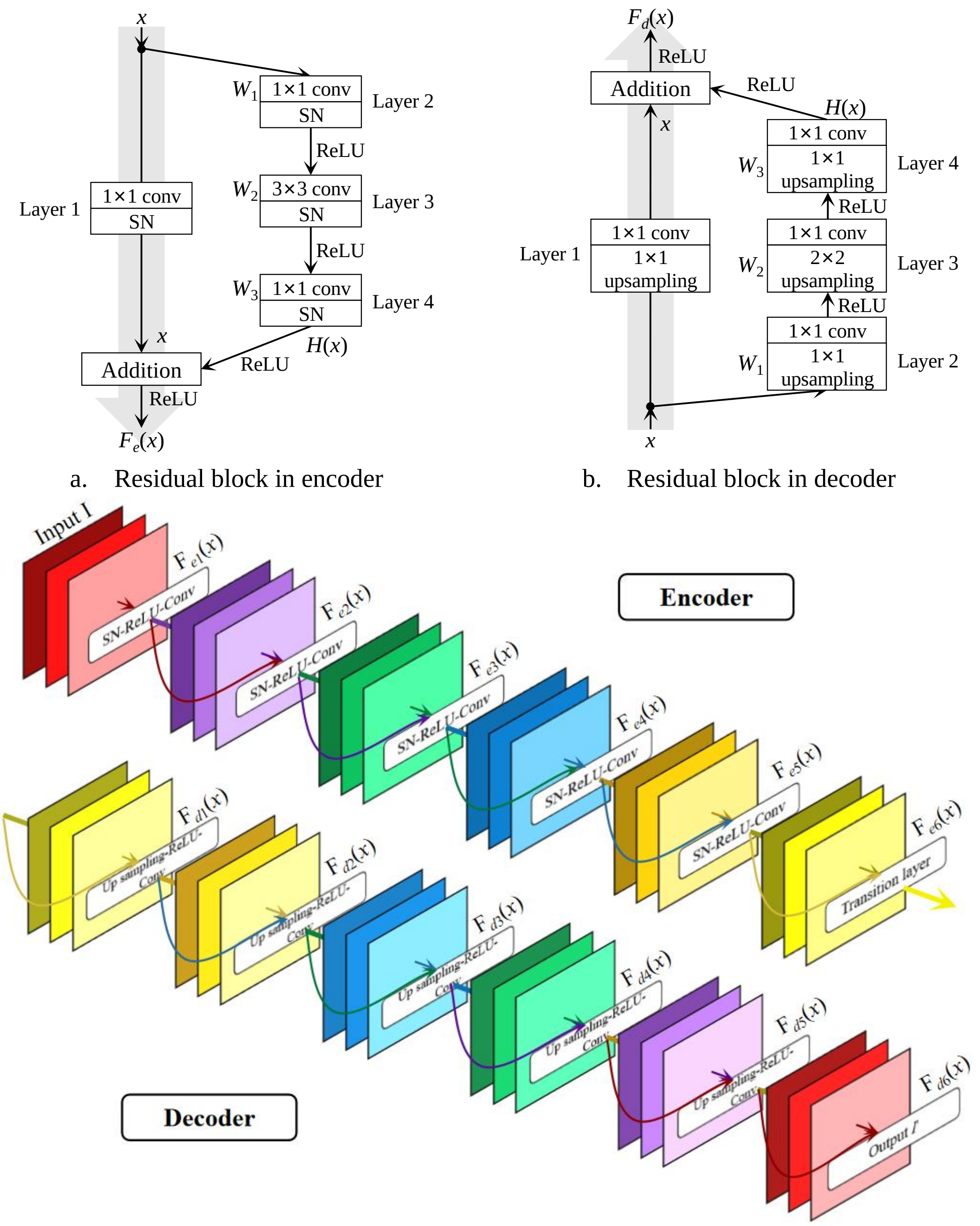

a. Residual block in encoder

b. Residual block in decoder

c. The ResVAE

**Fig. 3.** The architecture of the ResVAE.

If the Residual block without shortcut connection (the layer 1 from *x* to *x*), it is a typical three-layer-feedforward network. The training process is to let the output $F(x)=H(x)$ be $x$. While, as for the Residual block, the optimization is to let the $F(x)=H(x)-x$ be 0, which is much easier.

Moreover, normalization techniques are effective components in deep learning to avoid gradient disappear and accelerate training speed. In recent years, many normalization methods, such as Batch Normalization (BN) [28], Instance Normalization (IN) [29], and Layer Normalization (LN) [30], have been developed. Despite their great successes, existing works often employ the same normalizer in all normalization layers, rendering suboptimal performance [31]. Thus, Luo [31] addressed a learning-to-normalize problem by proposing Switchable Normalization (SN), which was learnt to select different normalizers for different normalization layers of a DPNN. The SN employs three distinct scopes to compute statistics (means and variances) including a channel, a layer and a minibatch, and switches between them by learning their importance weights in an end-to-end manner. The SN can maintain high-performance even when the batch size is too small, thus the computational requirement can be greatly reduced. Therefore, an SN (the detailed normalization can refer to the Appendix B) layer is added to each convolutional layer of the block. The detailed Residual block is shown in Table 1. Through the blocks in the encoder, the length and width of the tensor reduce 2 times while the depth increases 2 times. As for the decoder, it is a reverse process of the encoder, and the up sampling process mainly through the Bilinear Interpolation as shown in Appendix C.

**Table 1** Detailed architecture of the Residual block.

| | Layer | Input size | Output size | Kernel size | Stride | Padding |
|---|---|---|---|---|---|---|
| Encoder | Layer 1 | $[h, w, c]$ | $[0.5h, 0.5w, 2c]$ | [1, 1] | 1 | SAME |
| | Layer 2 | $[h, w, c]$ | $[h, w, c]$ | [1, 1] | 2 | SAME |
| | Layer 3 | $[h, w, c]$ | $[0.5h, 0.5w, c]$ | [3, 3] | 1 | SAME |
| | Layer 4 | $[0.5h, 0.5w, c]$ | $[0.5h, 0.5w, 2c]$ | [1, 1] | 2 | SAME |
| Decoder | Layer 1 | $[h, w, c]$ | $[2h, 2w, 0.5c]$ | [1, 1] | 1 | SAME |
| | Layer 2 | $[h, w, c]$ | $[h, w, c]$ | [1, 1] | 1 | SAME |
| | Layer 3 | $[h, w, c]$ | $[2h, 2w, 0.5c]$ | [1, 1] | 1 | SAME |
| | Layer 4 | $[2h, 2w, 0.5c]$ | $[2h, 2w, 0.5c]$ | [1, 1] | 1 | SAME |

The ResVAE is shown in Figs. 3 (c). Input images $I$ are compressed as 256-dimensional matrix $\mathbf{z}_{256}(z_1, z_2, \ldots, z_{256})$ through 6 convolutional Residual blocks. Such that, by inputting random $\mathbf{z}_{256}$ to the decoder, new physical cloud image $I'$ is easily obtained.

$$I_i' \leftarrow g\left(\mathbf{z}_i\left(z_1, z_2, \ldots\ldots, z_{256}\right)\right) \tag{4}$$

### 2.1.2. Prediction model

As discussed in Section 1, the physical cloud images contain more abundant, objective and graphic information and might be more suitable to build the surrogate models compared with limited design parameters. Thus, as shown in Fig. 4, different from traditional surrogate model-based optimizations, an intermediate variable, physical cloud image $I$ from Finite Element Analysis (FEA), is added between input parameters and actual responses. Figure 4 (a) is the traditional surrogate model that constructs a mapping from input parameters $\boldsymbol{\alpha}$ to actual responses $\hat{f}(\boldsymbol{\alpha})$. Then the EAs can handle the mapping relationship to achieve the optimum. While in Fig. 4 (b), an image $I$ is added. Through DPNN-based surrogate models, the surjection relationships between FE image, design parameters and actual responses can be established, where the DPNN is applied to do the dimension reduction. Thus, as shown in Fig. 1, the expensive-simulation is saved by using lower-cost generation and prediction models.

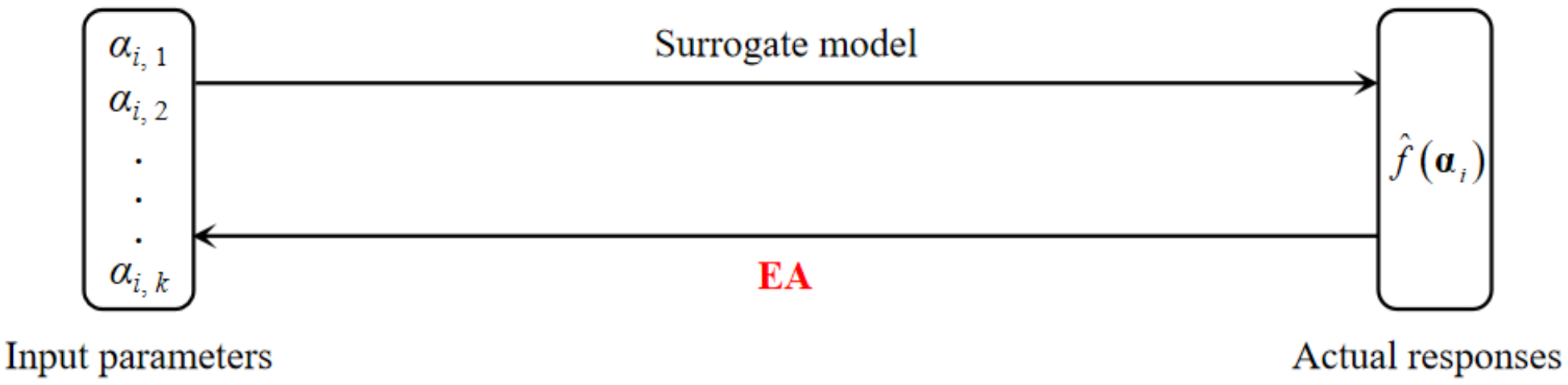

b. Traditional surrogate model-based optimization

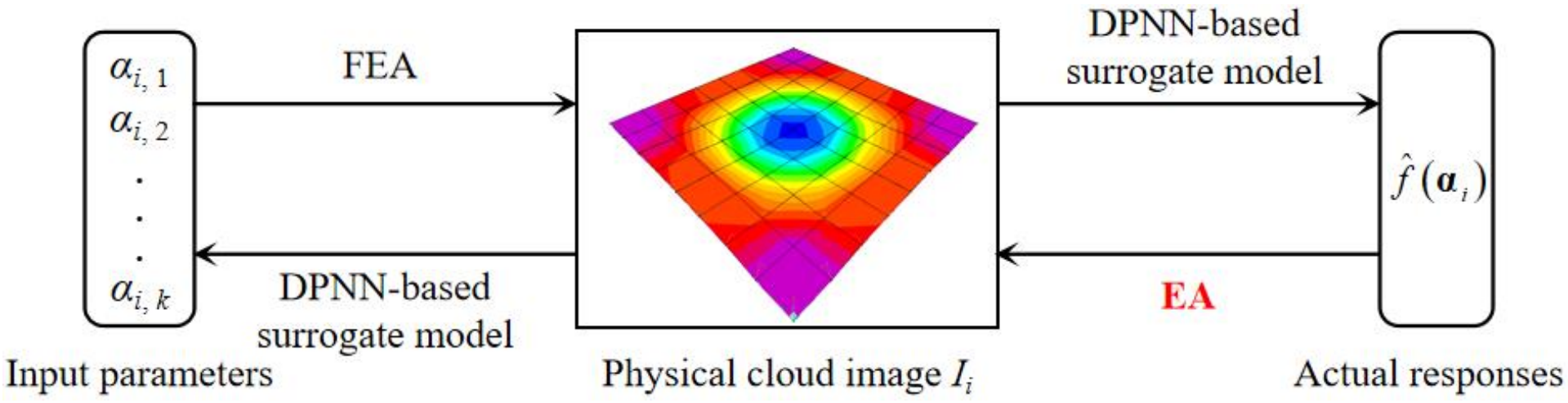

c. GRN-based optimization

**Fig. 4.** Comparison between GRN-based and surrogate model-based optimizations.

As for the dimensionality reduction, high-dimensional data, meaning data requiring more than two or three dimensions to be represented, is difficult to be interpreted. One approach to simplification is to assume that the interested data lie on an embedded nonlinear manifold within the lower-dimensional space. If the manifold is of low enough dimension, the data can be visualized in the low-dimensional space [18]. Hence, as a manifold learning model, another ResVAE reduces the generated image *I'* to **f**. Thus, the relationships from physical cloud images to input parameters and objective functions, which are constructed by simulations, are established by two surrogate models *h* (·), respectively, as

$$\boldsymbol{\alpha}_i \leftarrow h_1\left(\mathbf{f}_i\right)+\delta_1 \tag{5}$$

$$f\left(\boldsymbol{\alpha}_i\right) \leftarrow h_2\left(\mathbf{f}_i\right)+\delta_2 \tag{6}$$

where, $\delta_1$ and $\delta_2$ include both approximation and random errors.

In this study, the LSSVR is employed to construct the surrogate mappings from **f** to objection functions and design parameters, respectively. Here, **f** is designed as 16-dimensional.

2.1.3. New case filter

As shown in Fig. 1, there is a case filer between the generation and prediction models. That is because the generation model ResVAE, as shown in Appendix A, lets the prior over the latent variables be the centered isotropic multivariate Gaussian $p_\theta(\mathbf{z}|\mathrm{x})=N(\mathbf{z};\ \mathbf{0};\ I)$. Therefore, after training, the latent variables $p_\theta(\mathbf{z}|\mathrm{x})$ are satisfied with standard normal distribution. As shown in Fig. 5, 400 sheet forming samples in Section 3.2 are used to train the generation model. It can be seen all latent variables satisfy the standard normal distribution. Thus, only if the input to the generation model is a normal distribution $p(\mathbf{z})\sim N(0,\ 1)$, scientific and reasonable case can be generated. However, in EAs, a randomness is utilized to ensure the evolution of the algorithm. For example, in the PSO, two independent random sequences, $r_1 \sim U\ (0,\ 1)$ and $r_2 \sim U\ (0,\ 1)$, affect the stochastic nature. Hence, either initial input $p_\theta(\mathbf{z})$ is a normal or a random distribution, there will be non-normal distributions inputting to the generation model, which results in unreasonable new cases as shown in Fig. 6. These cases lack convincing details and suffer blurred regions, which make them neither realistic enough nor have sufficiently high resolution, especially those regions marked by red. Thus, these cases cannot be considered in the optimization and should be filtered out.

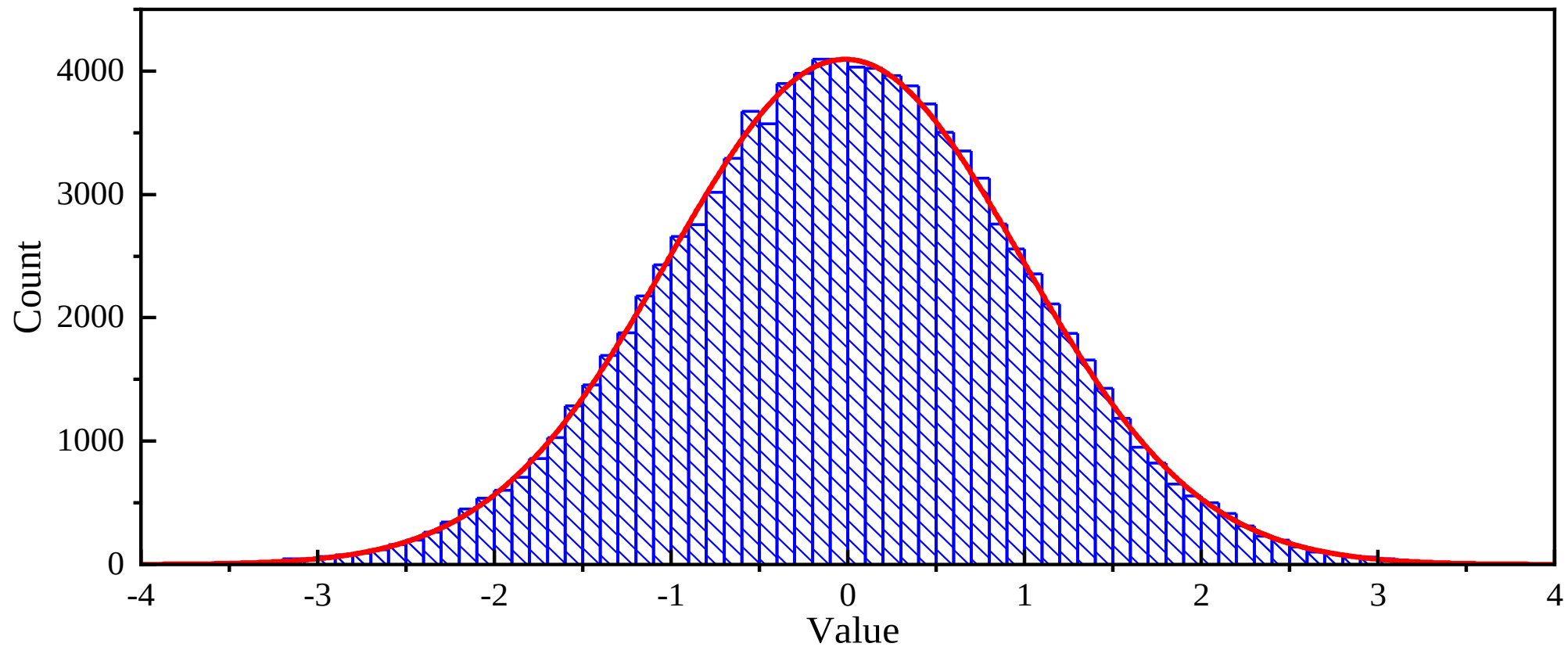

**Fig. 5**. Standard normal distribution of the latent variables.

Accordingly, between generation and prediction models, it needs a filter to ensure the generated cases considered in the EA are reasonable. Here, the image processing technique is applied. As shown in Fig. 7, an $m \times n \times 3$ HSV (Hue, Saturation, Value) image $H_o$ is simulated as the standard. Through image processing, if a pixel is white ([(0, 180), (0, 30), (221, 255)]), the pixel value is changed as (0, 0, 255) (white). Otherwise, the pixel is changed as black (0, 0, 0) (the range of black is [(0, 180), (0, 255), (0, 46)]). In this way, the outlined image $U_o$ can be drawn. Similarly, obtain the outlined image $U_i$ of each generated case $H_i$ and draw

$$J_i = abs\left(U_i - U_0\right) \tag{7}$$

The regions with the same color are (0, 0, 0) (black) in the $J_i$, and the ones with different colors are (0, 0, 255) (white). The white parts in $J_i$ are regarded as noises. According to Eq. (8) and through testing, if $n_{noise} > C$[1], the image $H_i$ is considered as unreasonable and should be filtered.

[1] $C$ is a preset constant. For different cases, it is different. In this study, the $C$ is 3,600 and 1,000 for the problems in Sections 3.1 and 3.2, respectively.

$$n_{noise} = \frac{1}{255}\sum_{h=1}^{H}\sum_{w=1}^{W}\sum_{c=1}^{3} J_i\left(h,w,c\right) \tag{8}$$

Ultimately, the architecture of the proposed GRN is illustrated in Fig. 8, and the sheet forming problem in Section 3.2 is diagrammed. To improve the universality of the algorithm, all input images are resized to 256×256×3 by the Bilinear Interpolation. By the way, for convenience, the less important parts (encoder of the first ResVAE and decoder of the second ResVAE) of the GRN are blurred.

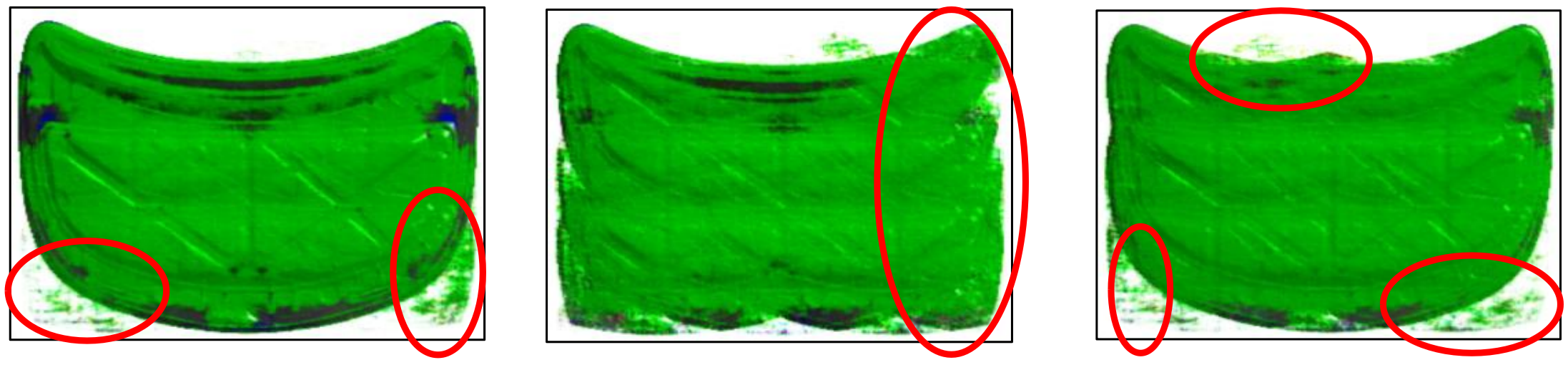

**Fig. 6.** Unreasonable examples.

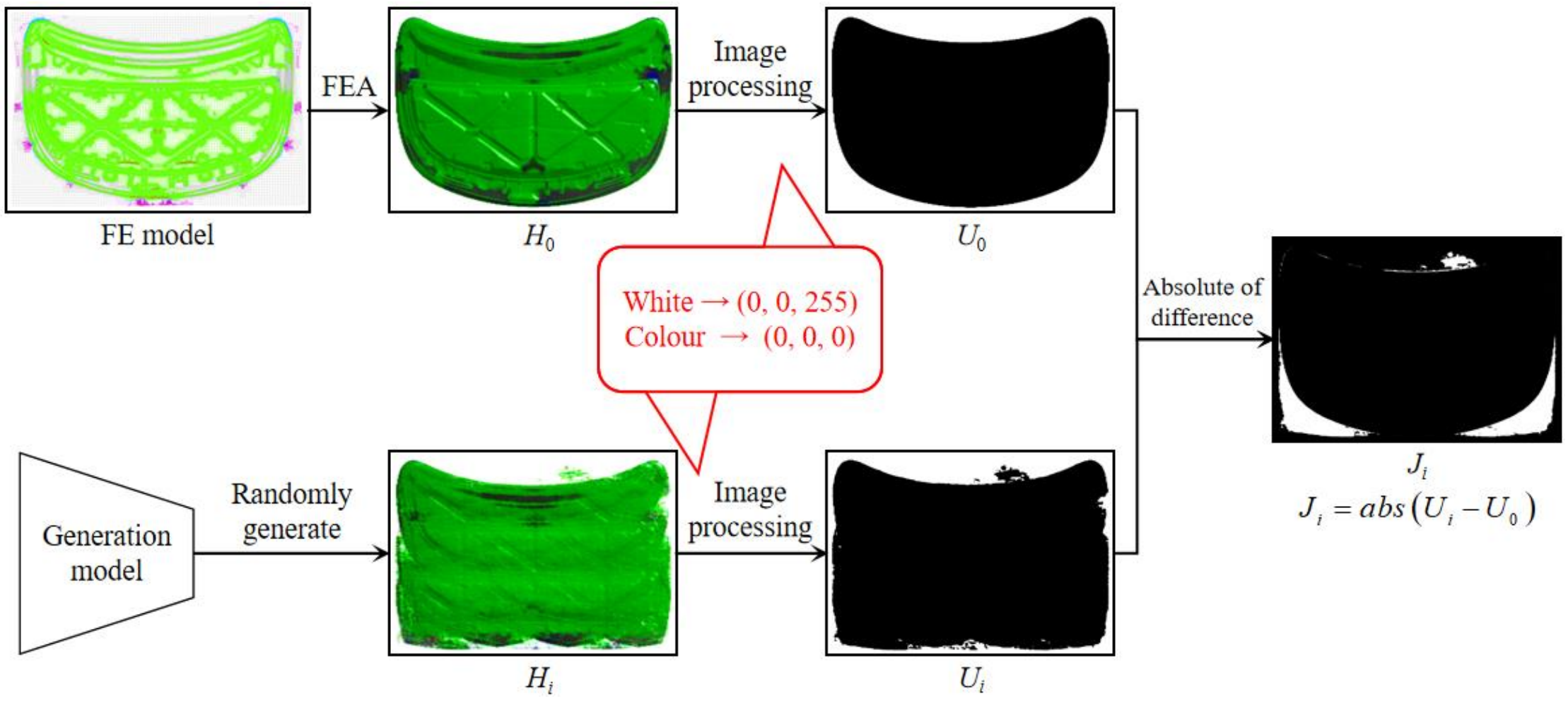

**Fig. 7.** The filter process of new cases.

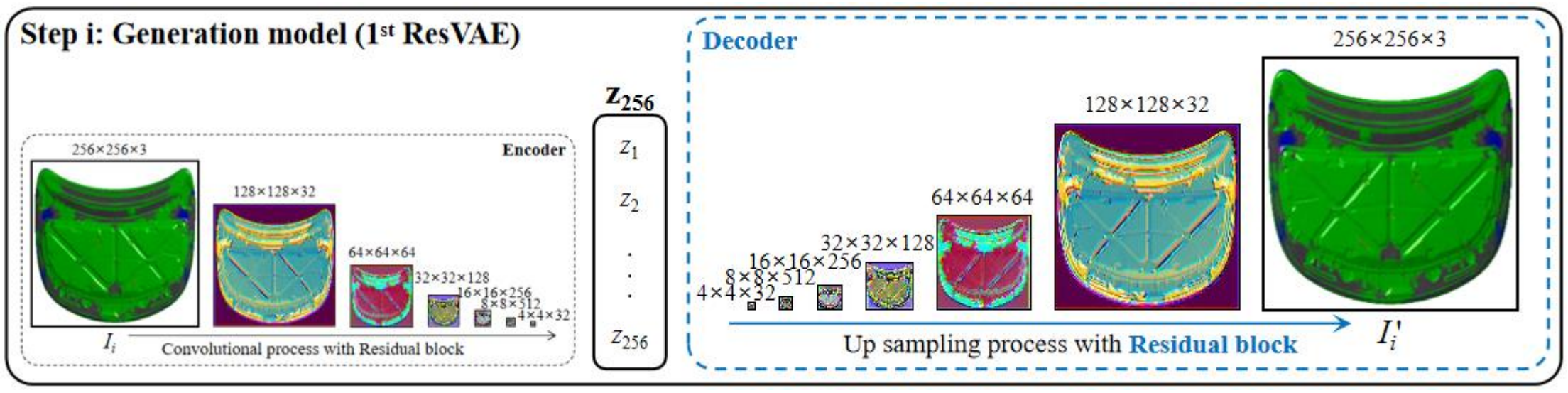

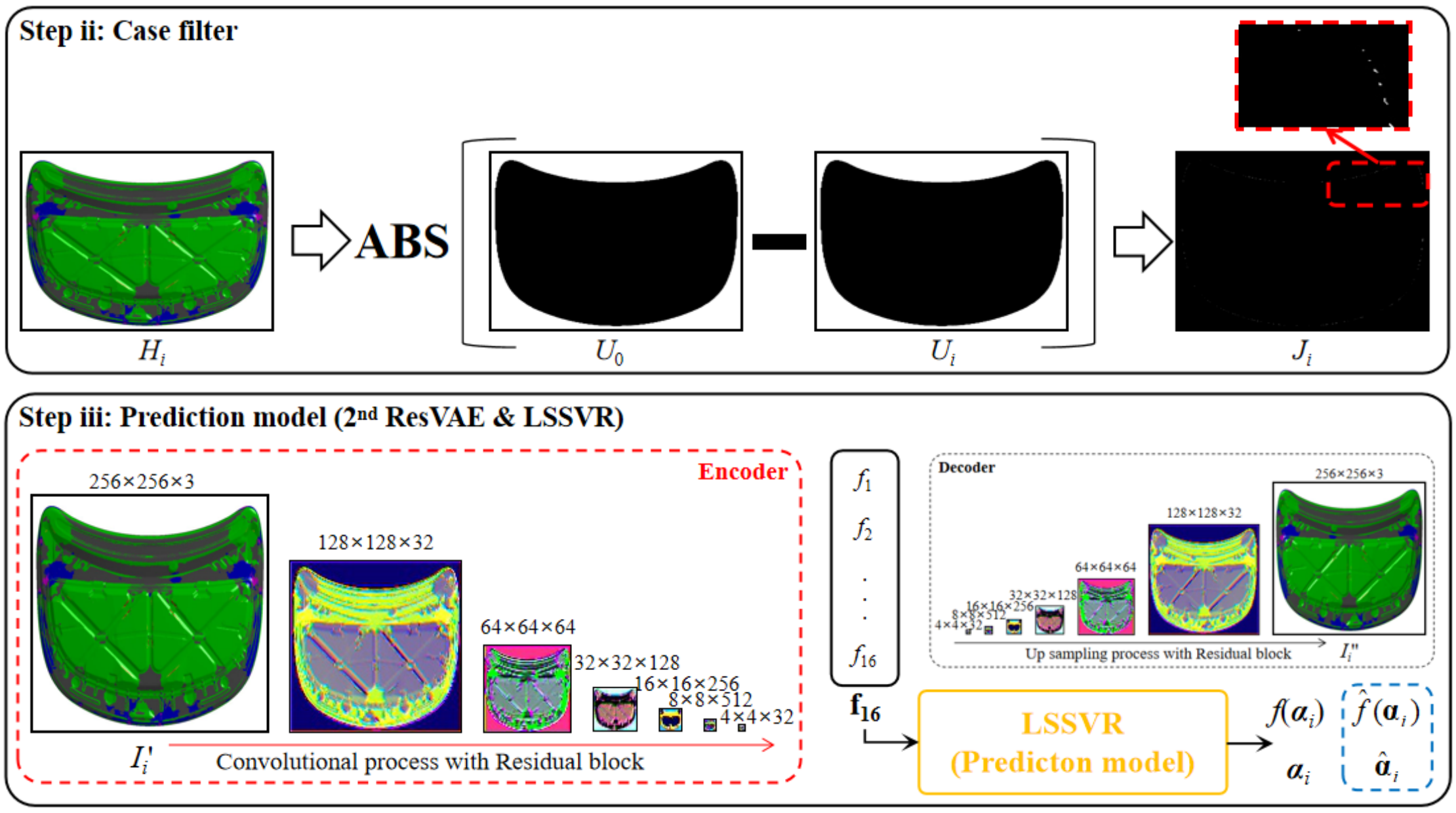

**Fig. 8**. Schematic of the GRN

## 2.2. Added-Randomized Particle Swarm Optimizer

As shown in Fig. 1, after constructing the GRN, EA [32] is employed to achieve the optimum. The optimization task is to determine a set of parameters, so that some measures of optimality, subject to certain constraints, are satisfied. As for a maximization/minimization task, the EA is stated as

Given $f$

$$\max/\min f = f\left(\mathbf{y}^j\left(y_1, y_2, \ldots, y_i\right)\right) \tag{9}$$

s.t.

$$\mathbf{y}^j \in \mathbb{S}^n \tag{10}$$

where $f(\cdot)$ is a complex problem; and $\mathbf{y}^j$ is a set of parameters belonging to the design space $\mathbb{S}^n$.

The EA mainly contains two steps. The 1st step accounts for the selection of representative points in the decision space, and the optimum is selected via crossover, mutation and selection. While the 2nd step is to determine the actual response in the objective space. Since there are trade-offs among objective functions, the optimization generally has a set of optimum solutions in Pareto sense, namely there is no optimum that is more superior than the other designs in all objectives [33]. Currently, there are several EAs be available for obtaining the Pareto frontier, e.g., Genetic Algorithm (GA), Particle Swarm Optimizer (PSO), and Ant Colony Optimization (ACO). Because of some of the attractive characteristics of the PSO including the ease of implementation and no gradient information being required, the PSO is selected. Actually, each EA method can be used here and it is not limited to the PSO.

The PSO is a population-based optimization method [34, 35]. It maintains a population of particles, where each particle represents a potential solution to an optimization problem. In this study, to help the PSO avoid the latent problem of the local optimum, random and partial particles are reset their positions to random positions periodically [36]. Thus, with the convergence of the PSO, the optimization process can satisfy the following Eq. (11) rule [37]. It means that when the optimization iteration approaches infinity, the PSO can ensure that all points with positive measure in the design space can be scanned and tested.

$$\prod_{k=0}^{\infty}\left(1-\mu_k\left[\mathbf{x}\right]\right)=0 \tag{11}$$

where $\mu_k[\mathbf{x}]$ is the probability of point **x** being generated by $\mu_k$.

Furthermore, considering the biggest difficulty in this study is the expensive simulation, the predetermined swarm size is largely reduced as 10. Then, another set of 4 random particles are added to the swarm in each step of the first 10 PSO iterations. Thus, the reduced swarm can help save lots of the simulation and time in the earlier stage. Meanwhile, the large enough swarm can be ensured for the final optimization. Moreover, the added particles can increase the reset particles in the start of the optimization, which can better satisfy the Eq. (11) rule. The resulting algorithm is called the Added-Randomized Particle Swarm Optimizer (ARPSO) and is represented as

**Algorithm** ARPSO.

**Require:** *step*, optimization iterations. $m_{\mathrm{add}}$, added swarm size in the first 10 iterations. *gbest*, the obtained optimum until current iterations. $\mathbf{x}[x_1, \ldots, x_m]$, particle set. *f*, a complex process.

1. **for** *s* in *step* **do**
2.     **for** $x_i$ in **x** **do**
3.         **if** $f(x_i) > gbest$

$gbest = f(x_i)$
**end if**
**end for**
Perform ARPSO updates.
**for** $j$ in int(10%$m$) **do**
$\mathbf{x}'[j] = \hat{x}$ where $\mathbf{x}'$ is disrupted $\mathbf{x}$ and $\hat{x}$ is a new random particle
**end for**
**if** $s \leq 10$
**for** $k$ in $m_{\text{add}}$ **do**
Add $\hat{x}$ to $\mathbf{x}$
**end for**
**end if**
**end for**

## 3. Numerical test and analyses

To evaluate the proposed GRN-EA method, a numerical example, a variable-stiffness composite hole-plate, is optimized in this section to determine the optimal distribution of curved fiber. Subsequently, a more complex engineering application, a sheet forming problem, is employed to test the proposed method for practical issues.

### 3.1. Numerical example: variable-stiffness composite hole-plate

Currently, composite materials play an increasingly important role due to its high-strength, high-stiffness and light-weight. Therefore, investigations into composite materials attract extensive attention. Compared to traditional straight-line Fiber Reinforcement Composite (FRC), a curved fiber distribution can lead to variability in the stiffness. The curvilinear composite laminate is known as a Variable-Stiffness (VS) composite. Due to the variability of the fiber angle orientation, structures are more designable, but the design difficulty is simultaneously improved [38]. Therefore, this section aims at the fiber angle deviation of the VS composite to achieve the best distribution.

As shown in Fig. 9, an 8-ply VS composite hole-plate is introduced. Here, 400 cases are sampled by Latin Hypercube Sampling (LHS). The thickness of each ply is 0.5 *mm*. The left side is fixed and a 2 *N*/*m* force is enforced on the right side horizontally. The hole-plate is modeled by 560 shell quadrilateral elements and the Degree of Freedoms (DOFs) are 3744. Furthermore, the determination of fiber path $z(x, y)$ is a quadratic polynomial function as

$$z(x, y) = x + a_1 y + a_2 xy + a_3 x^2 + a_4 y^2 \tag{12}$$

The predefined parameters of the Eq. (12) of 8 plies are shown in Table 2. It can be seen that the $a_1$ and $a_2$ of adjacent plies are contrarily while $a_3$ and $a_4$ are the same, which make sure the adjacent plies are symmetric for the bi-stability problem. Thus, there actually are 16 design parameters. The design space of each parameter is $[0.5a_i, 1.5a_i]$, where $a_i$ ($i$ = 1, 2, 3, 4) is the predefined value in Table 2. For this case, the optimization objective is to minimize the maximum displacement along the $y$-direction.

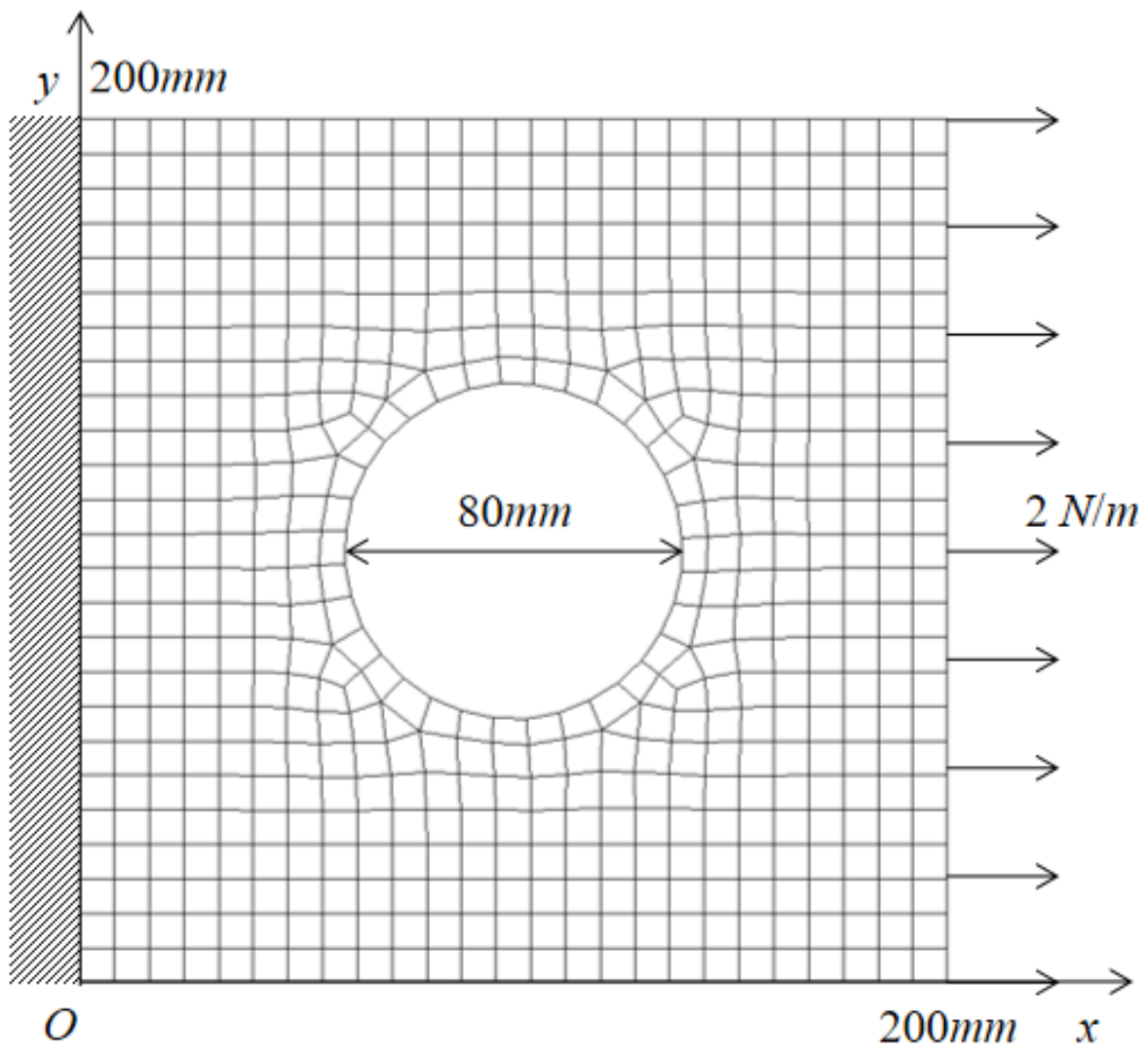

**Fig. 9.** The FE model of the VS composite hole-plate.

**Table 2** The predefined parameters.

| Ply | $a_1$ | $a_2$ | $a_3$ | $a_4$ |
|---|---|---|---|---|
| 1st ply | -2.879 | 0.527 | -0.015 | -9.989 |
| 2nd ply | 2.879 | -0.527 | -0.015 | -9.989 |
| 3rd ply | 6.270 | -0.656 | -1.745 | 7.811 |
| 4th ply | -6.270 | 0.656 | -1.745 | 7.811 |
| 5th ply | 7.258 | -0.069 | 4.087 | 17.191 |
| 6th ply | -7.258 | 0.069 | 4.087 | 17.191 |
| 7th ply | 14.277 | 2.586 | -2.127 | 10.519 |
| 8th ply | -14.277 | -2.586 | -2.127 | 10.519 |

### 3.1.1. Case generation and dimensionality reduction models.

Because both generation and dimensionality reduction models are the ResVAE, these two models are trained and evaluated together. As for the generation model, it needs to estimate the accuracy and reasonable of the generated cases (image *I*'). Similarly, for the dimensionality reduction model, the more accurate and reasonable of the restored images (image *I*''), the extracted $\mathbf{f}_{16}$ can represent more essential information. Consequently, Mean Square Error (MSE) is applied to show the error between generated and practical simulated cases.

$$MSE = \frac{1}{H \cdot W \cdot 3} \sum_{h}^{H} \sum_{w}^{W} \sum_{c}^{3} \left( I(h,\ w,\ c) - I'(h,\ w,\ c) \right)^2 \tag{13}$$

Additionally, Peak Signal-to-Noise Ratio (PSNR) [39] is employed to assess the image based on error sensitivity. While, Structural Similarity (SSIM) [40] judges the image rationality from brightness, contrast and structure. The PSNR is defined as

$$PSNR = 10 \cdot \log_{10}\left( \frac{p_{\max}^2}{MSE} \right) = 20 \cdot \log_{10}\left( p_{\max} \right) - 10 \cdot \log_{10}\left( MSE \right) \ (\mathrm{dB}) \tag{14}$$

where $p_{\max}$ is the maximum pixel value. Usually, it is 255.

The SSIM is a value between 0 and 1. The larger, the better. It is calculated by

$$SSIM = \frac{\left(2\mu_I \mu_{I'} + c_1\right)\left(2\sigma_{I,I'} + c_2\right)}{\left(\mu_I^2 + \mu_{I'}^2 + c_1\right)\left(\sigma_I^2 + \sigma_{I'}^2 + c_2\right)} \tag{15}$$

s.t.

$$c_1 = \left(k_1 L\right)^2, c_2 = \left(k_2 L\right)^2 \tag{16}$$

where $\mu_I$ and $\mu_{I'}$ are the pixel averages of *I* and *I*', respectively; $\sigma_I^2$ and $\sigma_{I'}^2$ are the pixel variances of *I* and *I*', respectively; $\sigma_{I,I'}$ is the covariance of *I* and *I*'; $c_1$ and $c_2$ stabilize the division with weak denominators; and *L* is the dynamic range of the pixel values. $L$=255, $k_1$=0.01 and $k_2$=0.03 by default.

**Table 3** Training results of ResVAEs.

| | PSNR | SSIM | MSE |
|---|---|---|---|
| Generation model | 32.70 | 0.959 | 5.39E-4 |
| Dimension reduction model | 40.77 | 0.985 | 8.49E-5 |

The training results (mean values) of the two ResVAEs, based on 400 samples, are shown in Table 3. Empirically, if the PSNR and SSIM are larger than 20 and 0.90, respectively, the ResVAE is satisfied. Moreover, the small enough MSE indicates that the generated VS composite hole-plates have high accuracy and the compressed low-dimensional $\mathbf{f}_{16}$ can well express the essential characteristics of generated cases.

Moreover, generation models are difficult to evaluate diversity. Therefore, two numerical assessment approaches, Inception Score (IS) [41] and Fréchet Inception Distance (FID) [42], for quantitative evaluation are attempted. The IS considers two aspects, one is intelligibility, and another is diversity. It employs the Inception Net-V3 [43] as an image classifier. Thus, if the intelligibility of an image $x$ is high enough, the classification result $p(y|x)$ should be a specific category, where $y$ is a 1,000-dimensional matrix that is output from Inception V3. While as for the diversity, the IS thinks that once the generated results are good-diversity, the generated images should uniform distribute in different categories, as follows $p(y)$ is uniform distribution. Mathematically, the IS is calculated as

$$IS = \exp\left(\mathbb{E}_x D_{KL}\left(p(y|x) \,\|\, p(y)\right)\right) \tag{17}$$

$$IS = \exp\left(\int_x D_{KL}\left(p(y|x) \,\|\, p(y)\right) dx\right) \tag{18}$$

where

$$D_{KL}\left(p(y|x) \,\|\, p(y)\right) = \sum_i p(y|x) \log \frac{p(y|x)}{p(y)} \tag{19}$$

Nevertheless, infinitesimal calculus cannot be calculated. Therefore, Eq. (19) is solved by inverse summation operation.

$$IS = \exp\left(\frac{1}{n}\sum_i^n p(y|x) \log \frac{p(y|x)}{p(y)}\right) \tag{20}$$

where $n$ is the number of the generated cases.

Thus, the larger the KL divergence, the better the diversity. However, the Inception Net-V3 is trained by the ImageNet [44]. The ImageNet is an image database organized according to the WordNet hierarchy, and it mainly contains nouns, e.g., dogs, fishes, birds, flowers, etc. Unfortunately, the dataset does not contain any simulated images. In other words, no matter the objective function of the generated hole-plate is large or small, and no matter the generated case is reasonable or unreasonable, almost all cases are categorized into the same category. That is

$$p(y_i | x) \approx p(y_i) \approx C, \; i = 0, \, 1, \, \ldots \tag{21}$$

Substitute Eq. (21) to Eq. (20),

$$\begin{aligned} IS &= \exp\left(\frac{1}{n}\sum_{i}^{n} p(y_i | x) \log \frac{p(y_i | x)}{p(y_i)}\right) \\ &\approx \exp\left(\frac{1}{n}\sum_{i}^{n} C \log \frac{C}{C}\right) = 1 \end{aligned} \tag{22}$$

It can be seen that the IS with 1 might be obtained for any generated model because the differences between training dataset (ImageNet) of the Inception V3 and hole-plate samples in this study are too large. Through calculations, the mean and standard deviation of the IS for the hole-plate generation model is 1.06 and 0.0069, which further authenticates the ratiocination. Similarly, the FID also uses the Inception Net as the classifier, and it faces the same issue as the IS. Thus, IS and FID cannot evaluate the model diversity in this study.

To sum up, the existing IS and FID are designed for the generation models of classification problems. In this study, the assessment ideas of the IS and FID are referenced and changed as that the normalized objective functions $f'(y|x)$ of generated cases should be uniformly distributed in the range [0, 1], where $f'(\cdot)$ indicates the normalized objective functions. Namely, if the generation model is diversity enough, the distribution of the normalized objective functions should be

$$\mathbb{E}_x\left[f'(y | x)\right] \rightarrow 0.5, \; \mathrm{var}\left[f'(y | x)\right] \rightarrow \frac{1}{12} \tag{23}$$

Here, the Eq. (23) is called as Case-Diversity Rule (CDR). Through 4,000 generated cases, the mean and variance values of the generated case are 0.61 and 0.0793, respectively, which conforms to the CDR.

#### 3.1.2. Prediction model

For the evaluation of regression problems, $R$ square ($R^2$) is usually used and calculated by

$$R^2 = 1 - \frac{\sum_i^n \left( f(\boldsymbol{\alpha}_i) - \hat{f}(\boldsymbol{\alpha}_i) \right)^2}{\sum_i^n \left( f(\boldsymbol{\alpha}_i) - \bar{f}(\boldsymbol{\alpha}) \right)^2} \quad (24)$$

where $\bar{f}(\boldsymbol{\alpha})$ is the mean of actual responses, and $n$ is the sample size.

Furthermore, Relative Average Absolute Error (RAAE) and Relative Maximum Absolute Error (RMAE) [10] are also employed to validate the approximation models. Similar to the $R^2$, the RAAE shows the overall accuracy of an approximation model. The closer the RAAE approaches zero, the more accurate. Besides, RMAE describes the error in a subregion of the design space, and a small RMAE is preferred. They are represented by

$$\mathrm{RAAE} = \sum_i^n \frac{\left| f(\boldsymbol{\alpha}_i) - \hat{f}(\boldsymbol{\alpha}_i) \right|}{n \cdot \mathrm{STD}} \quad (25)$$

$$\mathrm{RMAE} = \max\left(\frac{\left| f(\boldsymbol{\alpha}_i) - \hat{f}(\boldsymbol{\alpha}_i) \right|}{\mathrm{STD}}\right) \quad (26)$$

where STD stands for standard deviation of actual responses.

Here, two surrogate models are attempted to predict the objective functions (the displacement along the $y$-direction) and reverse the design parameters. One is BPNN, the most classical neural network structure, and another is LSSVR, a typical and traditional surrogate model. Here, 300 samples are for training and another 100 samples are for testing. The input is the 16-dimensional features $\mathbf{f}_{16}(f_1, f_2, \ldots, f_{16})$ compressed by the dimensionality reduction model, and the outputs are the objective functions and design parameters, respectively. Through testing, because there are too many parameters in the BPNN to train, 300 samples easily result in over-fitting, where the training and testing $R^2$ are 0.9999 and 0.1301, respectively. On the other hand, the testing results of the LSSVR are shown in Table 4. The LSSVR uses the kernel to fit the inputs in a higher linear space to obtain a good result. In addition, compared with the BPNN, the LSSVR needs less training samples and can better avoid the over-fitting problem. Hence, the prediction models use the LSSVR.

**Table 4** Training results of LSSVRs.

| | $R^2$ | RAAE | RMAE |
|---|---|---|---|
| Maximum *y*-displacement | 0.9891 | 0.0784 | 0.5039 |
| Design parameters (mean of 16 parameters) | 0.9415 | 0.1023 | 0.6012 |

### 3.1.3. Optimization by EA

The optimization process is shown in Fig. 10, after 100 iterations, the optimization has converged well and the obtained optimum is satisfied. After that, the optimal design parameters are shown in Table 5. Based on these parameters, the corresponding simulated displacement-physical cloud image is shown in Fig. 11.

**Table 5** The optimal design parameters.

| Ply | $a_1$ | $a_2$ | $a_3$ | $a_4$ |
|---|---|---|---|---|
| 1st ply | -2.2940 | 0.6346 | -0.0097 | -13.5071 |
| 2nd ply | 2.2940 | -0.6346 | -0.0097 | -13.5071 |
| 3rd ply | 4.5110 | -0.5156 | -2.4234 | 8.4524 |
| 4th ply | -4.5110 | 0.5156 | -2.4234 | 8.4524 |
| 5th ply | 5.4429 | -0.0539 | 2.4090 | 23.3557 |
| 6th ply | -5.4429 | 0.0539 | 2.4090 | 23.3557 |
| 7th ply | 17.1130 | 1.9114 | -2.9691 | 6.5913 |
| 8th ply | -17.1130 | -1.9114 | -2.9691 | 6.5913 |

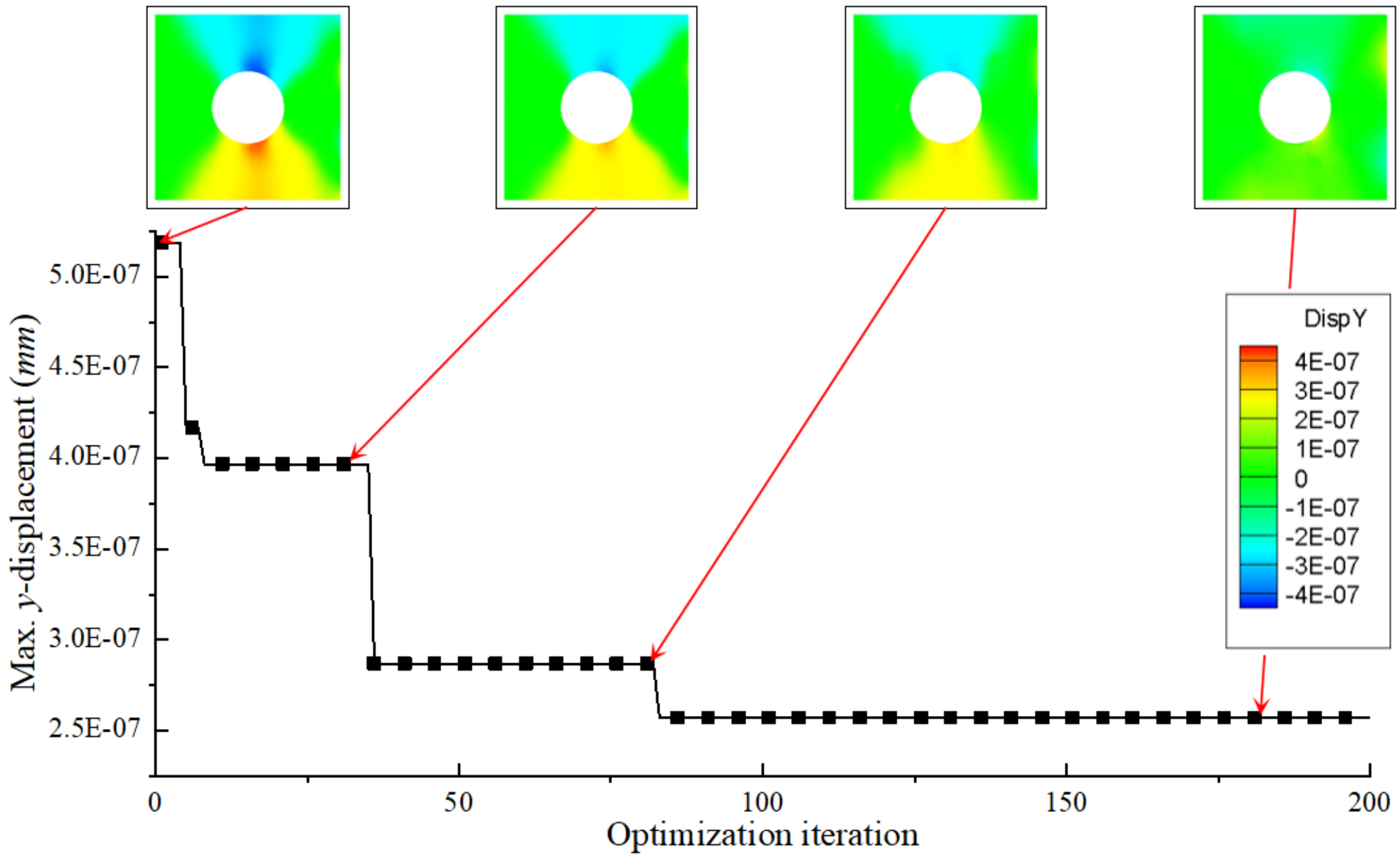

**Fig. 10.** Optimization processes by using the ARPSO.

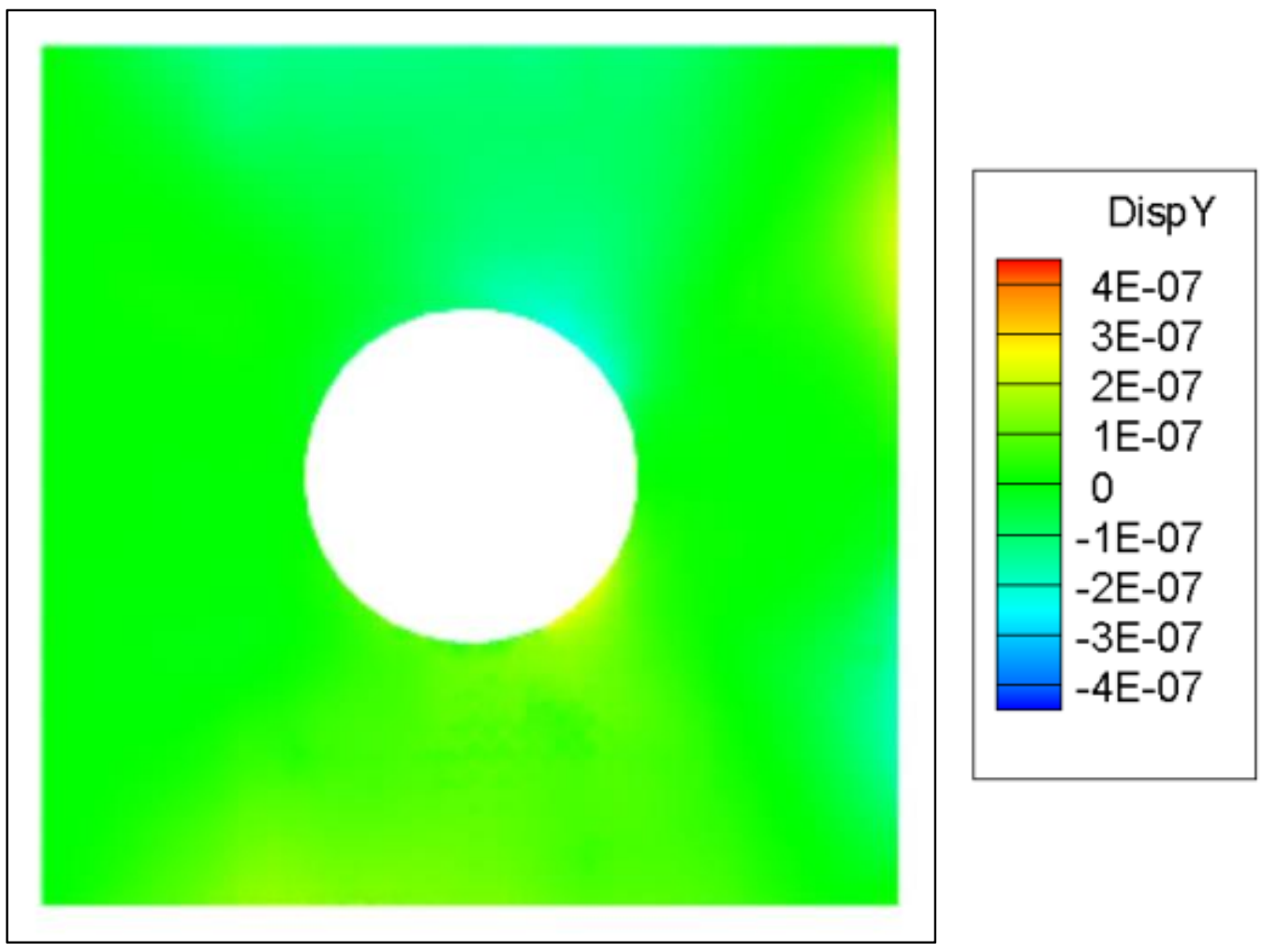

**Fig. 11.** The simulated displacement-physical cloud image of the optimal parameters.

### 3.2. Engineering application: sheet forming problem

After evaluation by a simple numerical example, the GRN-EA is introduced to a practical engineering application. With the development of the economy, production research and development (R&D) cycle in the automobile manufacturing industry keeps shortening. In the R&D cycle, sheet forming is an important factor. As the main panels of a car body, the designs of engine inner hoods are extremely significant. Therefore, the forming process of an engine inner hood is optimized by the proposed GRN-EA.

As shown in Fig. 12, LS-DYNA is used to simulate training samples. Here, 400 samples are achieved by LHS. Its Computer-Aided Design (CAD) model mainly consists of blank and tools, e.g., die, punch and binder. The thickness of the blank is 0.8 mm, and the material is DC04_0.80 mm (36). As shown in Fig. 12 (b), the blank is modeled by 8,066 quadrilateral elements, and the tools are modeled by 81,740 quadrilateral elements and 28,728 triangular elements. As shown in Table 6, images along with corresponding design parameters and objective functions, computed via nonlinear simulations, constitute the training dataset.

As for optimizations, objective functions should be given. The Forming Limit Diagram (FLD) [45] is a measure of sheet metal formability for investigation of cold forming properties of materials and is popularly used to estimate the forming behavior and provide a graphical description. The detailed description of the FLD is in Appendix D. Briefly, in the FLD, the more the green point percentage (safe regions) the better. However, there is a constraint condition that if a forming result contains any red points (risk of crack regions), this case is not considered no matter how many the green point percentage is. To sum up, the objective function is the green point percentage under the constraint of red point percentage that is represented by

**Require:** *r*, red point percentage. *g*, green point percentage.
1. **if** $r <$ E-8
2. $g = g$
3. **else**
4. $g = 0$
5. **end if**

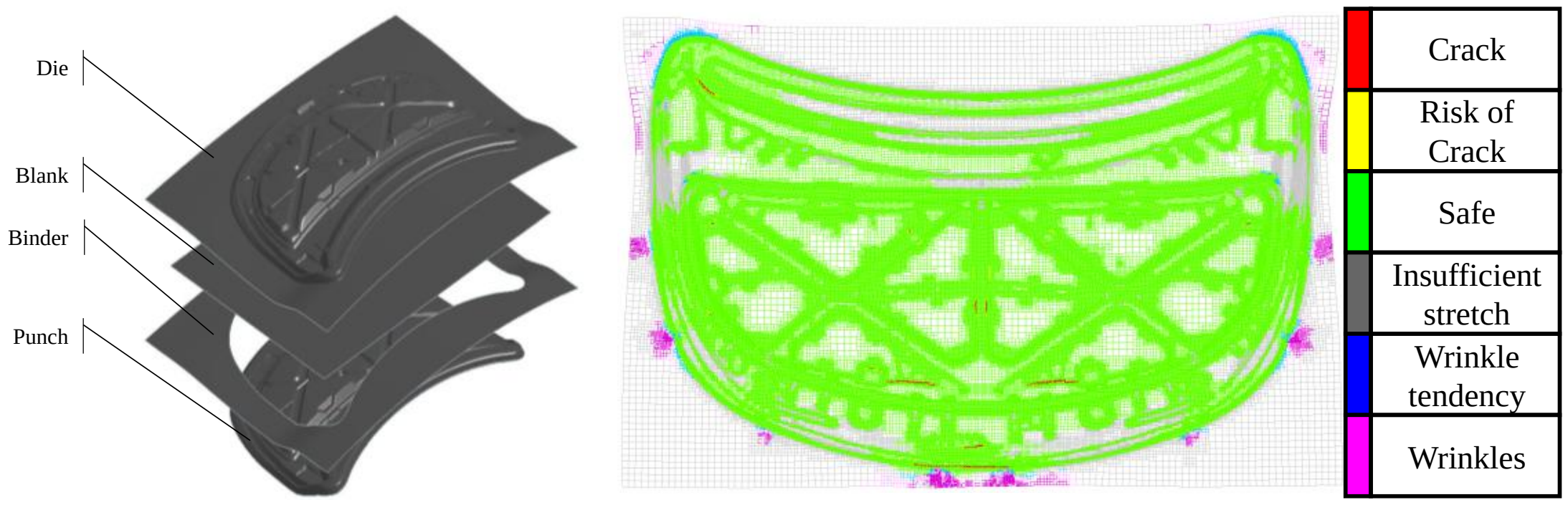

a. CAD model b. FE model

**Fig. 12.** The model of the engine inner hood.

**Table 6** The design parameters.

| | | | |
|---|---|---|---|
| Friction coefficient of punch/die | $f_1$ | Blank holder force (*Ton*) | BHF |
| Friction coefficient of binder | $f_2$ | Drawbead resistance (*N*/*mm*) | *F* |
| Drawing speed (*mm*/*s*) | *v* | | |

### 3.2.1. GRN construction.

Following the steps in Sections 3.1.1 and 3.1.2, two ResVAEs for generation and dimension-reduction are constructed. The satisfied training results (mean values of 400 samples) are shown in Table 7. Then, the modeling results of the prediction models, by the LSSVR, are shown in Table 8. The satisfied $R^2$ indicates the reasonable of the prediction models.

**Table 7** Training results of ResVAEs.

| | PSNR | SSIM | MSE |
|---|---|---|---|
| Generation model | 29.09 | 0.916 | 1.28E-3 |
| Dimension reduction model | 39.35 | 0.975 | 1.23E-4 |

**Table 8** Training results of LSSVRs.

| | Objective | $R^2$ | RAAE | RMAE |
|---|---|---|---|---|
| Objective functions | Red point percentage | 0.9029 | 0.2349 | 0.9910 |
| | Green point percentage | 0.9785 | 0.1074 | 0.5794 |
| Design parameters | Mean of 5 parameters | 0.9019 | 0.3105 | 1.0784 |

### 3.2.2. Optimization by EA

The optimization process is shown in Fig. 13. For 400 optimization iterations, the total time is about 56 *h*, compared with 50 × 400 × 2 *h* when optimizing the practical simulations, the costs is greatly and amazingly saved by using the GRN-EA. Furthermore, due to the case filter and the constraint of red points before the optimization, the methods can obtain a good result in the optimization start (green point percentage is bigger than 90%).

The optimal design parameters are shown in Table 9. Based on the result, the corresponding simulated FLD is shown in Fig. 14. Although the whole FLD is not all safe, the crack parts are avoided as much as possible. Therefore, the proposed algorithm can obtain the optimum with very low computational costs.

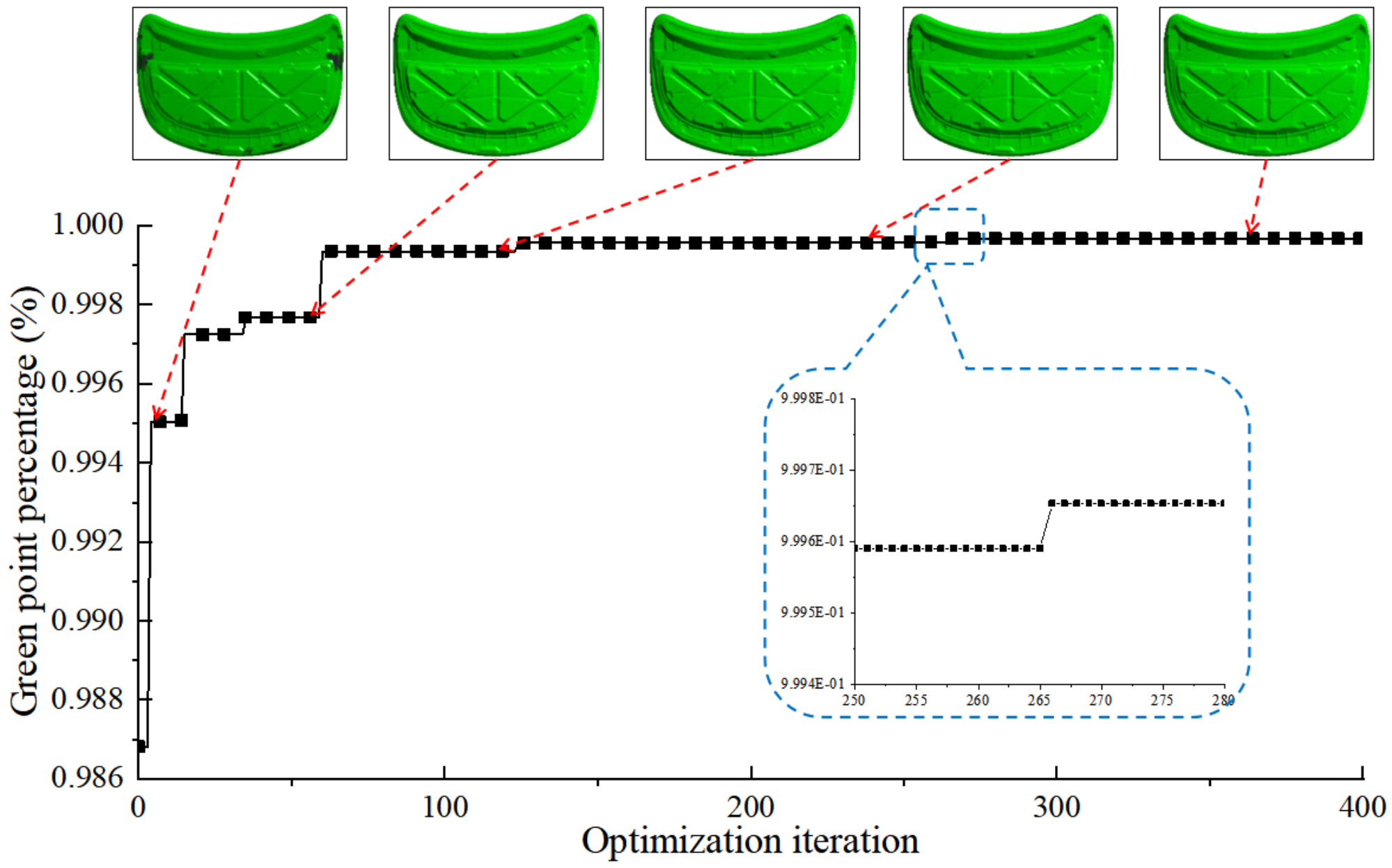

**Fig. 13.** Optimization processes of ARPSO.

**Table 9** The optimal design parameters.

| | | | |
|---|---|---|---|
| Friction coefficient of punch/die | $f_1$=0.1734 | Blank holder force (*Ton*) | 114.3358 |
| Friction coefficient of binder | $f_2$=0.1749 | Drawbead resistance (*N*/*mm*) | 110.4336 |
| Drawing speed (*mm*/*s*) | 3541.3534 | | |

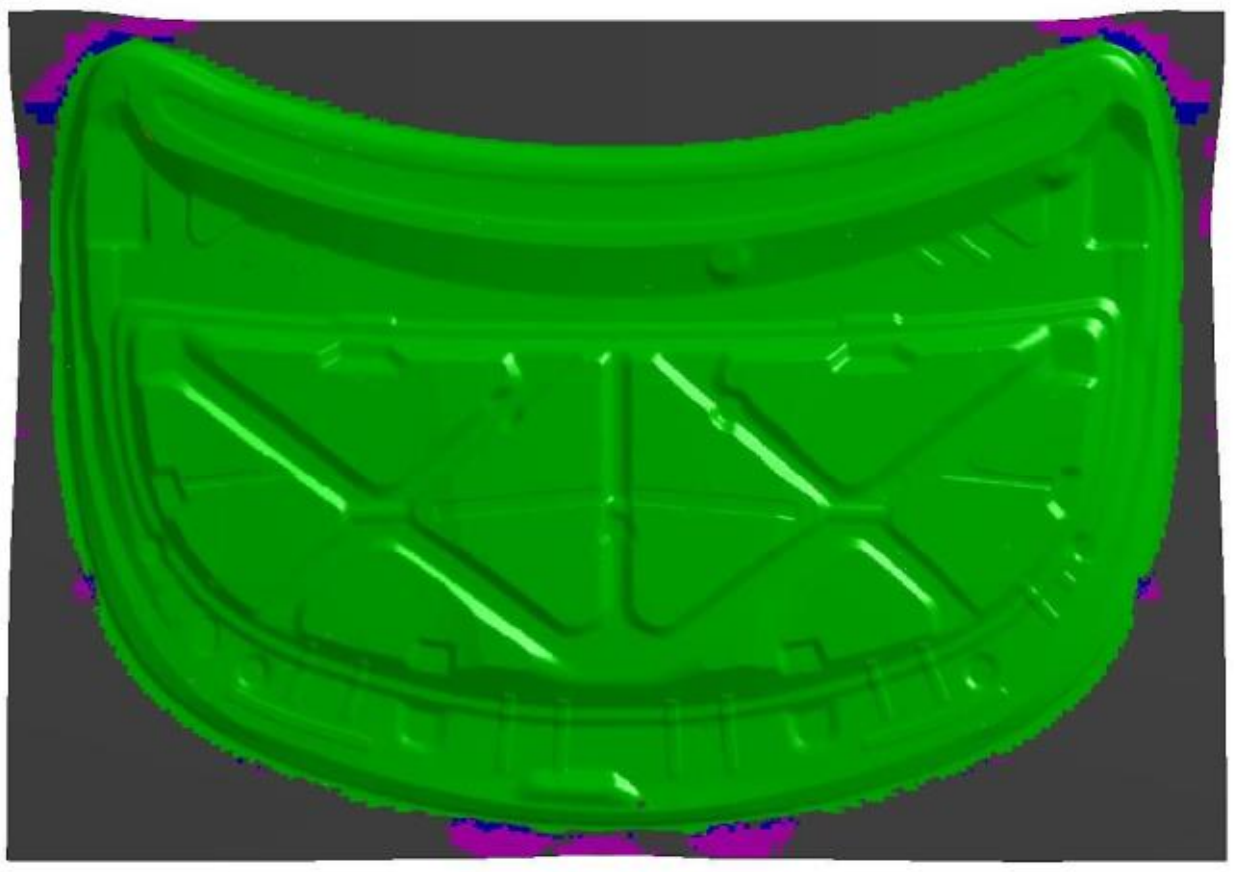

**Fig. 14.** The simulated FLD based on the achieved optimal design parameters.

## 4. Conclusions

In this study, a novel GRN-EA method is proposed and introduced to solve the bottleneck of the simulation-based optimization problems due the expensive-calculation costs. The contributions can be summarized as

i. Physical cloud images are attempted to construct the surrogate models instead of design parameters of the traditional surrogate models. Compared with limited designed parameters, cloud images contain more abundant, objective and graphic information, which can help them construct higher-accurate surrogate models.

ii. In order to avoid the gradient explosion or disappearance in deeper neural networks, the Residual block is introduced to the VAE. Thus, a generation model named as ResVAE is proposed.

iii. The inputs must be normal distributions when ResVAE is used as generation models. However, non-normal distributions must exist because of the randomness in the EA. Thus, a novel case filter by using image processing techniques is developed.

iv. For generation models, diversity must be evaluated. Nevertheless, existing evaluation methods, e.g. IS and FID, can only be used to the generation models for the classification problems, and there are no diversity criteria for the generation models of the regression problems. In this study, a CDR is proposed.

v. The proposed GRN-EA is successfully utilized a numerical example of a variable-stiffness composite hole-plate to obtained the optimal distribution of the curved fiber. After that, a more complex practical engineering problem, sheet-forming case, is also handled.

## Acknowledgments

This work has been supported by Project of the Key Program of National Natural Science Foundation of China under the Grant Numbers 11572120, 51621004 and 11972155, Key Projects of the Research Foundation of Education Bureau of Hunan Province (17A224).

## Appendix A. Variational Auto-Encoder

If there is a data set $\mathbf{x}[\mathrm{x}_i]$ and it is not big enough, the estimated distribution $p(\mathrm{x})$ is not accurate. So, an unobserved continuous random variable $\mathbf{z}$ is employed to assume that the data are generated based on the $\mathbf{z}$. Thus,

$$p(\mathrm{x}) = \int p(\mathrm{x} \mid \mathbf{z}) p_\theta(\mathbf{z}) d\mathbf{z} \tag{A1}$$

The true posterior density is

$$p_\theta(\mathbf{z} \mid \mathrm{x}) = \frac{p_\theta(\mathrm{x} \mid \mathbf{z}) p_\theta(\mathbf{z})}{p_\theta(\mathrm{x})} \tag{A2}$$

here, $\mathbf{z}$ is generated from some prior distribution $p_\theta(\mathbf{z})$; x is generated from some conditional distribution $p_\theta(\mathrm{x}|\mathbf{z})$.

It is different to evaluate or differentiate the marginal likelihood because $p_\theta(\mathrm{x})$ and $p_\theta(\mathbf{z}|\mathrm{x})$ by Eq. (A1) or Eq. (A2) cannot be calculated. Therefore, the VAE introduces a recognition model $q_\varphi(\mathbf{z}|\mathrm{x})$ as a probabilistic encoder, since given a data point x it produces a distribution $\mathbf{z}$ (e.g. a Gaussian or a uniform distribution). Similarly, $p_\theta(\mathrm{x}|\mathbf{z})$ is referred as a probabilistic decoder. Here, $\mathbf{z}$ is assumed to be $p(\mathbf{z}) \sim N(z, \mu, \sigma^2 \boldsymbol{I})$. After that, KL-divergence is employed to judge the differences between two distributions $q(\mathbf{z}|\mathrm{x})$ and $p(\mathrm{x}|\mathbf{z})$. It is calculated by

$$D_{KL}\left[q(\mathbf{z}\mid\mathrm{x})\| p(\mathbf{z}\mid\mathrm{x})\right]=\int_{q(\mathbf{z}|\mathrm{x})} q(\mathbf{z}\mid\mathrm{x})\log\frac{q(\mathbf{z}\mid\mathrm{x})}{p(\mathbf{z}\mid\mathrm{x})}$$

$$=\mathbb{E}_{q(\mathbf{z}|\mathrm{x})}\left[\log q(\mathbf{z}\mid\mathrm{x})-\log p(\mathbf{z}\mid\mathrm{x})\right]+\log p(\mathrm{x}) \tag{A3}$$

Define variational lower bound as

$$\mathfrak{L}(\theta,\varphi;\mathrm{x})=-\mathbb{E}_{q(\mathbf{z}|\mathrm{x})}\left[\log q(\mathbf{z}\mid\mathrm{x})-\log p(\mathbf{z}\mid\mathrm{x})\right] \tag{A4}$$

Because KL-divergence is non-negative,

$$\log p(\mathrm{x})\geq\mathfrak{L}(\theta,\varphi;\mathrm{x}) \tag{A5}$$

Hence, the purpose to differentiate and optimize the lower bound. Based on the maximum likelihood estimate,

$$\log p(\mathrm{x})=\sum_i \log p(x_i) \tag{A6}$$

where

$$\log p(x_i)=\mathfrak{L}(\theta,\varphi;x_i)+D_{KL}\left[q(\mathbf{z}\mid x_i)\| p(\mathbf{z}\mid x_i)\right] \tag{A7}$$

thus

$$\mathfrak{L}(\theta,\varphi;\mathrm{x})=-D_{KL}\left[q(\mathbf{z}\mid\mathrm{x})\| p(\mathbf{z})\right]+\mathbb{E}_{q(\mathbf{z}|\mathrm{x})}\left[\log p(\mathrm{x}\mid\mathbf{z})\right] \tag{A8}$$

As assumed above, $\mathbf{z}$ is a Gaussian distribution, so the KL-divergence can be written as

$$D_{KL}\left[q(\mathbf{z}\mid\mathrm{x})\| p(\mathbf{z})\right]=\int q(\mathbf{z}\mid\mathrm{x})\log\frac{q(\mathbf{z}\mid\mathrm{x})}{p(\mathbf{z})}$$

$$=\int q(\mathbf{z}\mid\mathrm{x})\log q(\mathbf{z}\mid\mathrm{x})d\mathbf{z}-\int q(\mathbf{z}\mid\mathrm{x})\log p(\mathbf{z})d\mathbf{z}$$

$$=\int N(\mathbf{z};\mu,\sigma^2)\log N(\mathbf{z};\mu,\sigma^2)dz-\int N(\mathbf{z};\mu,\sigma^2)\log N(\mathbf{z};0,I)dz$$

$$=-\frac{1}{2}\sum_j\left(1+\log(\sigma_j^2)-\mu_j^2-\sigma_j^2\right) \tag{A9}$$

As for the 2nd item of Eq. (A8), according to Gaussian distribution,

$$p(\mathrm{x}\mid\mathbf{z})=\frac{1}{\prod_i^D\sqrt{2\pi\sigma_i^2(\mathbf{z})}}e^{-\frac{1}{2}\left\|\frac{x-\mu(\mathbf{z})}{\sigma(\mathbf{z})}\right\|^2} \tag{A10}$$

$$-\ln p(\mathrm{x}\,|\,\mathbf{z}) = \frac{1}{2}\left\|\frac{x-\mu(\mathbf{z})}{\sigma(\mathbf{z})}\right\|^2 + \frac{D}{2}\ln 2\pi + \frac{1}{2}\sum_{i}^{D}\ln\sigma_i^2(\mathbf{z}) \tag{A11}$$

$$\ln q(\mathrm{x}\,|\,\mathbf{z}) - \ln p(\mathrm{x}\,|\,\mathbf{z}) = \frac{D}{2}\ln 2\pi + \frac{1}{2}\sum_{i}^{D}\ln\sigma_i^2(\mathbf{z}) \tag{A12}$$

s.t.

$$\ln q(\mathrm{x}\,|\,\mathbf{z}) \sim \frac{1}{2}\left\|\frac{x-\mu(\mathbf{z})}{\sigma(\mathbf{z})}\right\|^2 \tag{A13}$$

Accordingly, to minimize the Eq. (A13) is to minimize the MSE. Therefore, the loss function of the VAE is the weighted summation of the MSE and the KL divergence. Furthermore, the distribution of **z** is further designed as normal distribution, so as long as the input ~ $N(\mathbf{0}, \boldsymbol{I})$, new reasonable cases can be generated.

## Appendix B. Switchable Normalization

Normalization do the shift and scale before the data $x$ being input to the next network layer. Assumue that $x \in \mathbb{R}^{N\times C\times H\times W}$, namely the dataset contans $N$ samples (batch size), and the channel, height and width of each sample are $C$, $H$ and $W$, respectively.

i. Batch Normalization

The BN do the nomalization along the batch-direction as

$$\mu_{BN}(x) = \frac{1}{NHW}\sum_{n}^{N}\sum_{h}^{H}\sum_{w}^{W} x_{nchw} \tag{B1}$$

$$\sigma_{BN}^2(x) = \frac{1}{NHW}\sum_{n}^{N}\sum_{h}^{H}\sum_{w}^{W}\left(x_{nchw} - \mu_{BN}(x)\right)^2 \tag{B2}$$

The nomilized result is

$$\hat{x} = \gamma\frac{x-\mu_{BN}(x)}{\sqrt{\sigma_{BN}^2(x)+\varepsilon}} + \beta \tag{B3}$$

where $\varepsilon$ is a small positive value to avoid the denominator is zero, and $\gamma$ and $\beta$ are a scale and a shift parameter, respectively.

ii. Instance Normalization

Firstly, the IN do the nomalization along the chinnal-direction, and then along the batch-direction.

$$\mu_{IN}\left(x\right)=\frac{1}{HW}\sum_{h}^{H}\sum_{w}^{W}x_{nchw} \tag{B4}$$

$$\sigma_{IN}^{2}\left(x\right)=\frac{1}{HW}\sum_{h}^{H}\sum_{w}^{W}\left(x_{nchw}-\mu_{IN}\left(x\right)\right)^{2} \tag{B5}$$

The nomilized result is

$$\hat{x}=\gamma\frac{x-\mu_{IN}\left(x\right)}{\sqrt{\sigma_{IN}^{2}\left(x\right)+\varepsilon}}+\beta \tag{B6}$$

iii. Layer Normalizaiton

For the LN, it do the nomalization mainly along the dimensions of channel, height and width.

$$\mu_{LN}\left(x\right)=\frac{1}{CHW}\sum_{c}^{C}\sum_{h}^{H}\sum_{w}^{W}x_{nchw} \tag{B7}$$

$$\sigma_{LN}^{2}\left(x\right)=\frac{1}{CHW}\sum_{c}^{C}\sum_{h}^{H}\sum_{w}^{W}\left(x_{nchw}-\mu_{LN}\left(x\right)\right)^{2} \tag{B8}$$

The nomilized result is

$$\hat{x}=\gamma\frac{x-\mu_{LN}\left(x\right)}{\sqrt{\sigma_{LN}^{2}\left(x\right)+\varepsilon}}+\beta \tag{B9}$$

iv. Switchable Normalization

In the SN, adaptive normalization method is proposed to intergrete BN, IN and LN.

$$\mu_{SN}=\omega_{BN}\mu_{BN}+\omega_{IN}\mu_{IN}+\omega_{LN}\mu_{LN} \tag{B10}$$

$$\sigma_{SN}^{2}=\omega'_{BN}\sigma_{BN}^{2}+\omega'_{IN}\sigma_{IN}^{2}+\omega'_{LN}\sigma_{LN}^{2} \tag{B11}$$

s.t.

$$\omega_{k}=\frac{e^{\lambda_{k}}}{\sum_{z\in\{BN,IN,LN\}}e^{\lambda_{z}}},\ \omega'_{k}=\frac{e^{\lambda'_{k}}}{\sum_{z\in\{BN,IN,LN\}}e^{\lambda'_{z}}},\ k\in\{BN,\ IN,\ LN\} \tag{B12}$$

where each $\omega_k$ is computed by using a softmax function with $\lambda_{BN}$, $\lambda_{IN}$, and $\lambda_{LN}$ as the control parameters, which can be learned by Back Propagation (BP); and each $\omega'_k$ can be similarly achieved by $\lambda'_{BN}$, $\lambda'_{IN}$, and $\lambda'_{LN}$.

The nomilized result is

$$\hat{x} = \gamma \frac{x - \mu_{SN}(x)}{\sqrt{\sigma_{SN}^2(x) + \varepsilon}} + \beta \tag{B13}$$

## Appendix C. Bilinear Interpolation

Bilinear Interpolation is an extension of the linear interpolation for interpolating the functions of two variables (e.g., *x* and *y*) on a rectilinear 2D grid. The key idea is to perform linear interpolation first in one direction, and then again in another direction. Although each step is linear, the interpolation as a whole is not linear but rather quadratic.

If a point ($x$, $y$) of an unknown function $f(x)$ is interpolated through another four points $Q_{ii}=(x_i, y_i)$, $Q_{ij}=(x_i, y_j)$, $Q_{ji}=(x_j, y_i)$ and $Q_{jj}=(x_j, y_j)$, the linear interpolation in the *x*-direction is first done and this yields

$$f(x, y_i) \approx \frac{x_j - x}{x_j - x_i} f(Q_{ii}) + \frac{x - x_i}{x_j - x_i} f(Q_{ji}) \tag{C1}$$

$$f(x, y_j) \approx \frac{x_j - x}{x_j - x_i} f(Q_{ij}) + \frac{x - x_i}{x_j - x_i} f(Q_{jj}) \tag{C2}$$

After that, the interpolation in the *y*-direction is processed to obtain the desired estimate.

$$f(x, y) \approx \frac{1}{(x_j - x_i)(y_j - y_i)} \begin{bmatrix} x_j - x & x - x_i \end{bmatrix} \begin{bmatrix} f(Q_{ii}) & f(Q_{ij}) \\ f(Q_{ji}) & f(Q_{jj}) \end{bmatrix} \begin{bmatrix} y_j - y \\ y - y_i \end{bmatrix} \tag{C3}$$

## Appendix D. Forming Limit Diagram

The empirical formula of the FLD is concluded as

$$\begin{cases} \varepsilon_1 = fld_0 + \varepsilon_2 \cdot (4.2\varepsilon_2 - 0.627) \\ \varepsilon_1 = fld_0 + \varepsilon_2 \cdot (-0.86\varepsilon_2 - 0.785) \end{cases} \tag{D1}$$

s.t.

$$fld_0 = \frac{n \cdot \left(23.36 + 14.042 \cdot \min\left(t,\ 3.0\right)\right)}{0.2116 \cdot 100} \tag{D2}$$

where $\varepsilon_1$ is principal strain; $\varepsilon_2$ is minor strain; $fld_0$ is the limit strain point of the plane strain state; $n$ is the strain hardening exponent; and $t$ is sheet thickness.

The Wrinkling Limit Curve (WLC) approximates a straight line and can be represented as

$$\varepsilon_1 = -\varepsilon_2 \ \left(\varepsilon_2 \le 0\right) \tag{D3}$$

Through simulations of the sheet forming, the principal strain $\varepsilon_1^i$ and minor strain $\varepsilon_2^i$ of a node $i$ can be obtained. Based on the FLD and WLC, drawing breakage value $p_i$ of the node $i$ is calculated as

(1) $\varepsilon_2^i > 0$

$$\begin{cases} p_i = \varepsilon_1^i - \left[ fld_0 + \varepsilon_2^i \cdot \left(-0.86 \cdot \varepsilon_1^i + 0.785\right)\right],\ \varepsilon_1^i > fld_0 + \varepsilon_2^i \cdot \left(-0.86 \cdot \varepsilon_1^i + 0.785\right) \\ p_i = 0,\ \varepsilon_1^i \le fld_0 + \varepsilon_2^i \cdot \left(-0.86 \cdot \varepsilon_1^i + 0.785\right) \end{cases} \tag{D4}$$

(2) $\varepsilon_2^i \le 0$

$$\begin{cases} p_i = \varepsilon_1^i - \left[ fld_0 + \varepsilon_2^i \cdot \left(4.2 \cdot \varepsilon_1^i + 0.627\right)\right],\ \varepsilon_1^i > fld_0 + \varepsilon_2^i \cdot \left(4.2 \cdot \varepsilon_1^i + 0.627\right) \\ p_i = 0,\ \varepsilon_1^i \le fld_0 + \varepsilon_2^i \cdot \left(4.2 \cdot \varepsilon_1^i + 0.627\right) \end{cases} \tag{D5}$$

After sheet forming, the wrinkle value $q_i$ of the node $i$ is

$$q_i = \begin{cases} -\left(\varepsilon_2^i + \varepsilon_1^i\right),\ \varepsilon_1^i < -\varepsilon_2^i \\ 0,\ \varepsilon_1^i \ge -\varepsilon_2^i \end{cases} \tag{D6}$$

Thus, the drawing breakage value $y_p$ and wrinkle value $y_q$ of the sheet metal forming is

$$y_p = \sum_{i=1}^{m} p_i^2 \tag{D7}$$

$$y_q = \sum_{i=1}^{m} q_i^2 \tag{D8}$$

where $m$ is the number of nodes.